\documentclass[11pt]{article}
\usepackage{a4p,cite}
\newcommand{\rb}{\mbox{$R_{\rm b}$}}
\newcommand{\rc}{\mbox{$R_{\rm c}$}}
\newcommand{\zb}{\mbox{$\rm Z^0$}}
\newcommand{\udsbar}{\mbox{$\rm u\overline{u}+d\overline{d}+s\overline{s}$}}
\newcommand{\ccbar}{\mbox{$\rm c\overline{c}$}}
\newcommand{\bbbar}{\mbox{$\rm b\overline{b}$}}
\newcommand{\qqbar}{\mbox{$\rm q\overline{q}$}}
\newcommand{\ztobb}{\mbox{$\zb\rightarrow\bbbar$}}
\newcommand{\ztocc}{\mbox{$\zb\rightarrow\ccbar$}}
\newcommand{\ztohadrons}{\mbox{$\zb\rightarrow\rm hadrons$}}

\newcommand{\gbb}    {\Gamma_{\rm b\overline{b}}}
\newcommand{\ghad}      {\Gamma_{\rm had}}
\newcommand{\gbbghaf}{\frac{\gbb}{\ghad}}
\newcommand{\gbbghad}{\gbb/\ghad}
\newcommand{\epluseminus}{\mbox{$\rm e^+e^-$}}
\newcommand{\euds}{\mbox{$\epsilon^{\rm uds}$}}
\newcommand{\ec}{\mbox{$\epsilon^{\rm c}$}}
\newcommand{\eb}{\mbox{$\epsilon^{\rm b}$}}

\newcommand{\pb}{\mbox{$p_{\rm B}$}}
\newcommand{\pbbar}{\mbox{$p_{\rm\bar{B}}$}}
\newcommand{\nfrag}{\mbox{$N_{\rm frag}$}}
\newcommand{\nfragbar}{\mbox{$N_{\rm\overline{frag}}$}}
\newcommand{\mean}[1]{\langle{#1}\rangle}
\newcommand{\cb}{\mbox{$C^{\rm b}$}}
\newcommand{\cc}{\mbox{$C^{\rm c}$}}
\newcommand{\cuds}{\mbox{$C^{\rm uds}$}}

\newcommand{\cbgeom}{\mbox{$C_{\rm geom}^{\rm b}$}}
\newcommand{\cbgeod}{\mbox{$C_{\rm geom}^{\rm b, data}$}}
\newcommand{\cbgeomc}{\mbox{$C_{\rm geom}^{\rm b, MC}$}}
\newcommand{\nv}{\mbox{$N_{\rm v}$}}
\newcommand{\nb}{\mbox{$N_{\rm\overline{v}}$}}
\newcommand{\nvv}{\mbox{$N_{\rm vv}$}}
\newcommand{\nvb}{\mbox{$N_{\rm v\overline{v}}$}}
\newcommand{\nbb}{\mbox{$N_{\rm\overline{vv}}$}}
\newcommand{\nab}{\mbox{$N_{\rm a\overline{v}}$}}

\newcommand{\nl}{\mbox{$N_{\rm\ell}$}}
\newcommand{\na}{\mbox{$N_{\rm a}$}}
\newcommand{\naa}{\mbox{$N_{\rm aa}$}}

\newcommand{\nll}{\mbox{$N_{\rm\ell\ell}$}}

\newcommand{\nvl}{\mbox{$N_{\rm v\ell}$}}
\newcommand{\nbl}{\mbox{$N_{\rm \overline{v}\ell}$}}

\newcommand{\fv}{\mbox{$f_{\rm v}$}}
\newcommand{\fb}{\mbox{$f_{\rm\overline{v}}$}}
\newcommand{\nt}{\mbox{$N_{\rm t}$}}
\newcommand{\ntt}{\mbox{$N_{\rm tt}$}}
\newcommand{\nhad}{\mbox{$N_{\rm had}$}}

\newcommand{\meanxe}{\mbox{$\langle x_E\rangle$}}
\newcommand{\lsred}{$L_R/\sigma_{L_R}$}
\newcommand{\vbar}{$\rm\overline{V}$}

\newcommand{\phiv}{\mbox{$\phi_{vv}$}}
\newcommand{\cpb}{\mbox{$C_{\pb}$}}
\newcommand{\cbp}{\mbox{$C_{p}^{\rm b}$}}
\newcommand{\cbpp}{\mbox{$C_{p,\bar{p}}^{\rm b}$}}
\newcommand{\pv}{\mbox{$p_{\rm v}$}}
\newcommand{\pvbar}{\mbox{$p_{\rm\bar{v}}$}}
\newcommand{\xb}{\mbox{$x_{\rm B}$}}
\newcommand{\xv}{\mbox{$x_{\rm v}$}}
\newcommand{\xvbar}{\mbox{$x_{\rm v}$}}
\newcommand{\nprim}{\mbox{$N_{\rm prim}$}}
\newcommand{\nprimbar}{\mbox{$N_{\rm\overline{prim}}$}}
\newcommand{\cpv}{\mbox{$C_{\pv}$}}
\newcommand{\cbvv}{\mbox{$C_{p_{\rm v},p_{\bar{\rm v}}}^{\rm b}$}}
\newcommand{\eh}{\mbox{$E_{\rm h}$}}
\newcommand{\ehbar}{\mbox{$E_{\rm\bar{h}}$}}
\newcommand{\ph}{\mbox{$p_{\rm h}$}}
\newcommand{\phbar}{\mbox{$p_{\rm\bar{h}}$}}
\newcommand{\mh}{\mbox{$m_{\rm h}$}}
\newcommand{\mhbar}{\mbox{$m_{\rm\bar{h}}$}}
\newcommand{\cbep}{\mbox{$C_{E,p}^{\rm b}$}}
\newcommand{\dedx}{{\rm d}E/{\rm d}x}
\newcommand{\dxnorm}{\mbox{$N^{\sigma}_{{\rm d}E/{\rm d}x}$}}
\newcommand{\costhb}{\mbox{$\cos\theta_{\rm B}$}}
\newcommand{\costhbbar}{\mbox{$\cos\theta_{\rm\bar{B}}$}}
\newcommand{\phib}{\mbox{$\phi_{\rm B}$}}
\newcommand{\phibbar}{\mbox{$\phi_{\rm\bar{B}}$}}
\newcommand{\PLB}[3] {Phys.~Lett.\ {B#1} (#2) #3}
\newcommand{\PRL}[3] {Phys.~Rev.\ {Lett.~#1} (#2) #3}
\newcommand{\PRD}[3] {Phys.~Rev.\ {D#1} (#2) #3}
\newcommand{\NIM}[3] {Nucl.~Instrum.\ {Methods~#1} (#2) #3}
\newcommand{\NPB}[3] {Nucl.~Phys.\ {B#1} (#2) #3}
\newcommand{\CPC}[3] {Comp.~Phys.\ {Comm.~#1} (#2) #3}
\newcommand{\ZPC}[3] {Z.~Phys.\ {C#1} (#2) #3}
\newcommand{\JPH}[3] {J.~Phys.\ {#1} (#2) #3}
\newcommand{\EPJ}[3] {Eur.~Phys.\ J.\ {C#1} (#2) #3}
\newcommand{\etal} {et~al.}
\input epsf
\newcommand{\epostfig}[3]{
\begin{figure}[tbp]
\setlength{\epsfxsize}{1.1\hsize}
\hspace*{-0.05\hsize} \epsfbox{#1}
\caption{\label{#2}#3}
\end{figure}
}
\newcommand{\cutv}{2.8}

\newcommand{\ntothad}{1\,923\,240}
\newcommand{\rbval}{0.2178}
\newcommand{\rbstat}{0.0011}
\newcommand{\rbsyst}{0.0013}
\newcommand{\rbsysq}{0.00129}
\newcommand{\delrbrc}{-0.056}
\newcommand{\delrbec}{-0.059}
\newcommand{\delrbeuds}{-0.010}
\newcommand{\rbtheo}{0.2155}
\newcommand{\rbterr}{0.0003}
\newcommand{\systrkres}{0.00017}
\newcommand{\systrkefi}{0.00014}
\newcommand{\syssidrop}{0.00009}
\newcommand{\syssialgn}{0.00008}
\newcommand{\syselidef}{0.00015}
\newcommand{\sysmuidef}{0.00009}
\newcommand{\syscqfrag}{0.00028}
\newcommand{\syschprov}{0.00031}
\newcommand{\syschprol}{0.00015}
\newcommand{\syschprod}{0.00046}
\newcommand{\syschlife}{0.00007}
\newcommand{\syschmult}{0.00014}
\newcommand{\syschneut}{0.00030}
\newcommand{\syschklam}{0.00015}
\newcommand{\syscsemil}{0.00031}
\newcommand{\syscsdmod}{0.00029}
\newcommand{\sysgsplcc}{0.00018}
\newcommand{\sysgsplbb}{0.00027}
\newcommand{\sysklhypr}{0.00001}
\newcommand{\sysmcstat}{0.00010}
\newcommand{\sysefitot}{0.00090}
\newcommand{\syselidbg}{0.00039}
\newcommand{\sysmuidbg}{0.00041}
\newcommand{\syscrltot}{0.00066}
\newcommand{\sysevtsel}{0.00033}
\newcommand{\elefierr}{4\,\%}

\begin{document}
\begin{titlepage}
{\center\Large

EUROPEAN LABORATORY FOR PARTICLE PHYSICS \\

}
\bigskip

{\flushright
CERN-EP/98-137 \\
August 31, 1998 \\
}
\begin{center}
    \LARGE\bf\boldmath
    A Measurement of \rb\ \\
    using a Double Tagging Method
\end{center}
\vspace{1cm}
\bigskip

\begin{center}
\Large The OPAL Collaboration \\
\bigskip
\large


\end{center}
\vspace{1cm}

\bigskip
\begin{abstract}
The fraction of \ztobb\ events in hadronic \zb\ decays
has been measured by the OPAL experiment using the data collected
at LEP between 1992 and 1995. The \ztobb\
decays were tagged using displaced secondary vertices, and 
high momentum electrons and muons.
Systematic uncertainties were reduced by measuring the
b-tagging efficiency using a double  tagging technique. Efficiency 
correlations between opposite hemispheres of an event are small, and
are well understood through comparisons between real and simulated
data samples. A value of 
\[
\rb\ \equiv \frac{\sigma (\rm e^+e^- \rightarrow\ \bbbar )}
{\sigma (\rm e^+e^- \rightarrow\ hadrons)}
= \rbval \pm \rbstat \pm \rbsyst 
\]
was obtained, where the first error is statistical and the second
systematic. The uncertainty on \rc, the fraction of \ztocc\ events
in hadronic \zb\ decays, is not included in the errors. The dependence
on \rc\ is
\[
\frac{\Delta\rb}{\rb} = \delrbrc \frac{\Delta\rc}{\rc} ,
\]
where $\Delta\rc$ is the deviation of \rc\ from the value 0.172
predicted by the Standard Model. The result for \rb\ agrees
with the value of $\rbtheo \pm\rbterr$ predicted by the Standard Model.
\end{abstract}

\vspace{1cm}     

\begin{center}
\large
Submitted to Eur.\ Phys.\ J.\ C.
%


\end{center}

\end{titlepage}
\begin{center}{\Large        The OPAL Collaboration
}\end{center}\bigskip
\begin{center}{
G.\thinspace Abbiendi$^{  2}$,
K.\thinspace Ackerstaff$^{  8}$,
G.\thinspace Alexander$^{ 23}$,
J.\thinspace Allison$^{ 16}$,
N.\thinspace Altekamp$^{  5}$,
K.J.\thinspace Anderson$^{  9}$,
S.\thinspace Anderson$^{ 12}$,
S.\thinspace Arcelli$^{ 17}$,
S.\thinspace Asai$^{ 24}$,
S.F.\thinspace Ashby$^{  1}$
% ...
}\end{center}\bigskip
\bigskip
$^{  1}$School of Physics and Astronomy, University of Birmingham,
Birmingham B15 2TT, UK
\newline
$^{  2}$Dipartimento di Fisica dell' Universit\`a di Bologna and INFN,
I-40126 Bologna, Italy
\newline
$^{  3}$Physikalisches Institut, Universit\"at Bonn,
D-53115 Bonn, Germany
\newline
$^{  4}$Department of Physics, University of California,
Riverside CA 92521, USA
\newline
$^{  5}$Cavendish Laboratory, Cambridge CB3 0HE, UK
\newline
$^{  6}$Ottawa-Carleton Institute for Physics,
Department of Physics, Carleton University,
Ottawa, Ontario K1S 5B6, Canada
\newline
$^{  7}$Centre for Research in Particle Physics,
Carleton University, Ottawa, Ontario K1S 5B6, Canada
\newline
$^{  8}$CERN, European Organisation for Particle Physics,
CH-1211 Geneva 23, Switzerland
\newline
$^{  9}$Enrico Fermi Institute and Department of Physics,
University of Chicago, Chicago IL 60637, USA
\newline
$^{ 10}$Fakult\"at f\"ur Physik, Albert Ludwigs Universit\"at,
D-79104 Freiburg, Germany
\newline
$^{ 11}$Physikalisches Institut, Universit\"at
Heidelberg, D-69120 Heidelberg, Germany
\newline
$^{ 12}$Indiana University, Department of Physics,
Swain Hall West 117, Bloomington IN 47405, USA
\newline
$^{ 13}$Queen Mary and Westfield College, University of London,
London E1 4NS, UK
\newline
$^{ 14}$Technische Hochschule Aachen, III Physikalisches Institut,
Sommerfeldstrasse 26-28, D-52056 Aachen, Germany
\newline
$^{ 15}$University College London, London WC1E 6BT, UK
\newline
$^{ 16}$Department of Physics, Schuster Laboratory, The University,
Manchester M13 9PL, UK
\newline
$^{ 17}$Department of Physics, University of Maryland,
College Park, MD 20742, USA
\newline
$^{ 18}$Laboratoire de Physique Nucl\'eaire, Universit\'e de Montr\'eal,
Montr\'eal, Quebec H3C 3J7, Canada
\newline
$^{ 19}$University of Oregon, Department of Physics, Eugene
OR 97403, USA
\newline
$^{ 20}$CLRC Rutherford Appleton Laboratory, Chilton,
Didcot, Oxfordshire OX11 0QX, UK
\newline
$^{ 22}$Department of Physics, Technion-Israel Institute of
Technology, Haifa 32000, Israel
\newline
$^{ 23}$Department of Physics and Astronomy, Tel Aviv University,
Tel Aviv 69978, Israel
\newline
$^{ 24}$International Centre for Elementary Particle Physics and
Department of Physics, University of Tokyo, Tokyo 113-0033, and
Kobe University, Kobe 657-8501, Japan
\newline
$^{ 25}$Institute of Physical and Environmental Sciences,
Brunel University, Uxbridge, Middlesex UB8 3PH, UK
\newline
$^{ 26}$Particle Physics Department, Weizmann Institute of Science,
Rehovot 76100, Israel
\newline
$^{ 27}$Universit\"at Hamburg/DESY, II Institut f\"ur Experimental
Physik, Notkestrasse 85, D-22607 Hamburg, Germany
\newline
$^{ 28}$University of Victoria, Department of Physics, P O Box 3055,
Victoria BC V8W 3P6, Canada
\newline
$^{ 29}$University of British Columbia, Department of Physics,
Vancouver BC V6T 1Z1, Canada
\newline
$^{ 30}$University of Alberta,  Department of Physics,
Edmonton AB T6G 2J1, Canada
\newline
$^{ 31}$Duke University, Dept of Physics,
Durham, NC 27708-0305, USA
\newline
$^{ 32}$Research Institute for Particle and Nuclear Physics,
H-1525 Budapest, P O  Box 49, Hungary
\newline
$^{ 33}$Institute of Nuclear Research,
H-4001 Debrecen, P O  Box 51, Hungary
\newline
$^{ 34}$Ludwigs-Maximilians-Universit\"at M\"unchen,
Sektion Physik, Am Coulombwall 1, D-85748 Garching, Germany
\newline
\bigskip\newline
$^{  a}$ and at TRIUMF, Vancouver, Canada V6T 2A3
\newline
$^{  b}$ and Royal Society University Research Fellow
\newline
$^{  c}$ and Institute of Nuclear Research, Debrecen, Hungary
\newline
$^{  d}$ and Department of Experimental Physics, Lajos Kossuth
University, Debrecen, Hungary
\newline
$^{  e}$ on leave of absence from the University of Freiburg
\newline
\newpage
\section{Introduction}
 
The partial width for the decay $\ztobb$ is of special interest in the
Standard Model.
Electroweak corrections involving the top quark affect the $\ztobb$
partial width, $\gbb$, differently from the widths for lighter quarks.
As a result, the fraction
\[
\gbbghaf \equiv \frac{\Gamma(\ztobb)}{\Gamma(\ztohadrons)}
\]
depends on the top quark mass, $m_{\rm top}$, but has negligible
uncertainty from the unknown Higgs boson mass and the strong coupling
constant $\alpha_s$.
The fraction $\gbbghad$ is also sensitive to various extensions
of the Standard Model
involving new particles such as additional quarks and gauge bosons,
or the virtual effects of new scalars and fermions
such as those expected in supersymmetric models~\cite{bamert}.

The quantity measured in this analysis is the cross-section ratio
\[
\rb\ \equiv \frac{\sigma (\rm e^+e^- \rightarrow\ \bbbar )}
{\sigma (\rm e^+e^- \rightarrow\ hadrons)}
\]
at the \zb\ resonance. This differs from the partial width ratio $\gbbghad$
because of the additional contribution from photon-exchange diagrams.
These have been evaluated within the Standard Model using the program
ZFITTER $5.0$ \cite{ZFITTER} which predicts that \rb\ is 0.0002 smaller
than $\gbbghad$. 
By convention,
\ztobb\ events where an additional \qqbar\ pair is produced via gluon
splitting are included in the numerators of the definitions of 
$\gbbghad$ and \rb . The small number of events where the only \bbbar\
pair is produced via gluon splitting,
rather than directly from \zb\ decay, is not included in the
numerators. The interference between these two processes, and the
effect on the measurement of \rb, are expected
to be negligible~\cite{theogcc}.

This paper supersedes the previously published
OPAL measurement~\cite{pr188} of the fraction of $\bbbar$ events
in hadronic $\zb$ decays. The measurement is improved by 
employing higher performance vertex tagging using the upgraded
silicon detector with two coordinate readout installed in 1993 
\cite{opalsi3d}, higher performance electron identification, and by
including the data taken in 1995. 

The paper is organised as follows. The analysis method, based on the 
double tagging technique, is described in the next section. The
OPAL detector, the selected event sample and the Monte Carlo
simulation are reviewed in Section~\ref{s:dsam}. The b-tagging methods
are discussed in Sections~\ref{s:vertex} and~\ref{s:lepton}.
The result is presented in Section~\ref{s:meas}, with systematic errors
being evaluated in Sections~\ref{s:syste} and~\ref{s:systc}.
A summary is given in Section~\ref{s:concl}.

\section{Analysis Method}\label{s:anna}

The analysis method is based on the double tagging
technique. Each selected hadronic \zb\ decay event is
divided into two hemispheres by the plane perpendicular to the thrust
axis and containing the interaction point. A b-tagging algorithm is
then applied separately to each hemisphere, and the number of tagged
hemispheres \nt\ and events where both hemispheres are tagged \ntt\
counted in the sample of \nhad\ selected hadronic events. These
quantities are related by:
\begin{eqnarray}
\nt & = & 2 \nhad \{ \eb\ \rb\ +\ec\ \rc\ +\euds\ (1-\rb - \rc ) \} ,
\label{e:ntsimp} \\
\ntt & = & \nhad \{ \cb\ (\eb )^2\ \rb +C^{\rm c} (\ec )^2\ \rc\ + 
C^{\rm uds} (\euds )^2\ (1-\rb -\rc )\} , \label{e:nttsimp}
\end{eqnarray}
where \eb , \ec\ and \euds\ are the tagging efficiencies for
hemispheres in \bbbar , \ccbar\ and light quark (uds) events,
and \cb, \cc\ and \cuds\ describe the tagging efficiency 
correlation between the two hemispheres in events of each flavour.
For a useful b-tagging algorithm, 
\eb\ is much larger than \ec\ and \euds .
The correlation \cb\ is defined by $\cb=\epsilon^{\rm bb}/(\eb)^2$,
where $\epsilon^{\rm bb}$ is the efficiency to tag both hemispheres of
a \bbbar\ event. Deviations of \cb\ from unity account for the fact that the
tagging in the two hemispheres is not completely independent, there
being a small efficiency correlation between them for both physical
and instrumental reasons. These correlations are also present in
\ccbar\ and light quark events; however their effect on the \rb\ measurement 
is less than $10^{-4}$ because the double tagging efficiencies for
these events are very small. The correlations \cc\ and \cuds\ are therefore
set to unity. The values of \rb\ and \eb\ can then be obtained by
solving equations~\ref{e:ntsimp} and~\ref{e:nttsimp}, and only the 
values of \ec, \euds\ and \cb\ need to be input from Monte Carlo simulation.
This technique avoids the need to input \eb\ from simulation, which 
severely limits the precision of single tag measurements of \rb.

In this analysis, two b-tags are used: a secondary vertex
tag with very high b-purity, 
and a high momentum lepton tag with somewhat lower b-purity and
efficiency. A hemisphere is taken to be tagged if it is tagged by
either one or both of the secondary vertex and lepton tags.
As the performance
of the tagging algorithms changes significantly between the different
years of data taking due to changes in the detector configuration,
the equations are solved separately for \eb\ and \rb\ for each
year. The values of \rb\ are then combined, taking into account
correlated and uncorrelated systematic uncertainties.
The charm tagging efficiencies \ec, light quark
tagging efficiencies \euds\ and \bbbar\
correlation coefficient \cb\ for each year are input from Monte Carlo
simulation, and are thus sources of systematic error. The relative
sizes of these systematic errors and the statistical error depend on
the purity and efficiency of the b-tag, and the tag cut is adjusted to
minimise the total resulting error. The only other
sources of systematic error are the value of \rc\ (which is fixed to
its Standard Model value), and the hadronic event selection.

\section{Data Sample and Event Simulation} \label{s:dsam}

The OPAL detector has been described in detail 
elsewhere~\cite{opalsi3d,opaldet}.
Tracking of charged particles is performed by a central detector,
consisting of a silicon microvertex detector, a vertex chamber, a jet chamber
and $z$-chambers\,\footnote{A right handed coordinate system is used, with
positive $z$ along the $\mathrm{e}^-$ beam direction and $x$ pointing
towards the centre of the LEP ring. The polar and azimuthal angles are
denoted by $\theta$ and $\phi$, and 
the origin is taken to be the centre of the detector.}.
The central detector is positioned inside a
solenoid, which provides a uniform axial magnetic field of 0.435\,T.
The silicon microvertex detector consists of two layers of
silicon strip detectors;
the inner layer covers a polar angle range of
$| \cos \theta | < 0.83$ and
the outer layer covers $| \cos \theta |< 0.77$.
This detector provided $\phi$ coordinate information in 1992, and
was upgraded to provide both $\phi$ and $z$ coordinate information from
1993. In 1995, a new detector geometry was installed with smaller gaps between 
the wafers in $\phi$, increasing the fraction of tracks with silicon hits
from both layers.
The vertex chamber is a precision drift chamber
which covers the range $|\cos \theta | < 0.95$.
The jet chamber is
a large-volume drift chamber, 4.0~m long and 3.7~m in diameter,
providing both tracking and ionisation energy loss (d$E$/d$x$) information.
The $z$-chambers provide a precise measurement of the $z$-coordinate
of tracks as they leave the jet chamber in the range
$|\cos \theta | < 0.72$.
The coil is surrounded by a
time-of-flight counter array and
a barrel lead-glass electromagnetic calorimeter with a presampler.
Including the endcap electromagnetic calorimeters,
the lead-glass blocks cover the range $| \cos \theta | < 0.98$.
The magnet return yoke is instrumented with streamer tubes
and serves as a hadron calorimeter.
Outside the hadron calorimeter are muon chambers, which
cover 93\% of the full solid angle.

The data used for this analysis were collected from \epluseminus\
collisions at LEP during 1992--1995, with centre-of-mass energies at
and around the peak of the \zb\ resonance. Hadronic events were selected
by the algorithm used in \cite{pr188}, giving a hadronic \zb\
selection efficiency of $(98.1\pm 0.5)\,\%$ and a background of less
than 0.1\%. The thrust value $T$ and thrust axis polar angle
$\theta_T$ were then calculated using charged tracks and 
electromagnetic clusters not associated to any tracks. To ensure a
well defined thrust axis direction within the acceptance of the
silicon microvertex detector, the thrust value and axis direction were required
to satisfy $T>0.8$ and $|\cos\theta_T|<0.7$. The complete selection has
an efficiency of about 58\,\% for hadronic \zb\ decays, and selected
a total of \ntothad\ events in the data.
Of these events, $4.9\,\%$ were
recorded below the peak of the \zb\ resonance at a centre-of-mass
energy of $E_{\rm cm}=89.4\rm\,GeV$ and $7.4\,\%$ above the peak at 
$E_{\rm cm}=93.0\rm\,GeV$. Calculations using ZFITTER \cite{ZFITTER}
indicate that these off-peak events change the measured value 
of \rb\ by $-0.00004$, a
correction which is $-0.02\,\%$ of the measured value and is 
not applied to the result presented here.

The event selection is designed to have the same efficiency for each
quark flavour. However, \bbbar\ events have a higher average
charged particle multiplicity than other flavours,
and hence a slightly higher probability of passing
the event selection requirement of at least
seven charged tracks. Owing to the high mass of the b quark, \bbbar\
events also have a slightly different thrust distribution,
but Monte Carlo studies showed this to have a 
much smaller effect than that
from the track multiplicity cut. According to the Monte Carlo,
these biases increase the measured 
value of \rb\ in the selected event sample by 0.32\,\% of its value. 
Non-hadronic background
(mainly $\rm e^+e^-\rightarrow\tau^+\tau^-$ events) constitutes about 
0.065\,\% of the selected events, and reduces the measured value of
\rb\ by 0.065\,\%. The combination of these event selection effects
increases the measured value of \rb\  by $(0.25\pm 0.15)\,\%$,
where the error is due to the
uncertainty in the simulation of the track multiplicity distributions
and  Monte Carlo statistics \cite{pr188}.

Charged tracks and electromagnetic calorimeter clusters with no
associated track were combined into jets using a cone algorithm 
\cite{jetcone} with a cone half angle of $R=0.55\rm\,rad$ and a 
minimum jet energy of $5\rm\,GeV$. This algorithm,
rather than the JADE algorithm employed in \cite{pr188}, increases the 
fraction of tracks in the jet coming from the b hadron decay,
and improves the b-tagging performance.

Monte Carlo simulated events were used for evaluating backgrounds,
acceptances for charm and light quark events, and efficiency correlations
between the two hemispheres of an event. Hadronic events were
simulated with the JETSET 7.4 generator \cite{jetset}, with parameters
tuned by OPAL \cite{jetsetopt}. The fragmentation function of Peterson et al.
\cite{fpeter} was used to describe the fragmentation of b and c quarks.
The generated events were passed through 
a program that simulated the response of the OPAL detector
\cite{gopal} and the same reconstruction algorithms as the data.
Further detailed studies of the tracking performance in the data were
made, and the results were used to tune the simulated tracking performance.


\section{Vertex Tagging}\label{s:vertex}

Hadronic \zb\ decays into b quarks are tagged by 
reconstructing secondary vertices significantly
separated from the primary vertex, taking
advantage of the relatively long ($\sim 1.5\rm\,ps$) lifetime, high
decay charged multiplicity and high mass ($\sim 5\rm\,GeV$) of 
b hadrons\,\footnote{The convention $c=1$ is assumed throughout this paper.}.
Information
characterising these features of b hadron production and decay 
is combined using an artificial neural network algorithm to produce a single
vertex tagging variable for each hemisphere.

\subsection{Primary Vertex Reconstruction}\label{ss:pvtx}

Primary vertices were reconstructed for each event
using the algorithm described in 
\cite{pr188}, but applied separately to each hemisphere of the event.
Reconstructing a separate independent primary vertex in each hemisphere avoids 
an efficiency correlation from sharing a single primary vertex
reconstructed using tracks from both hemispheres \cite{alephonetag}.
The position and uncertainty of the beam spot ($\rm e^+e^-$ collision
region) was used as a common 
constraint in both hemispheres, but its effect on the correlation is small.
The beam spot position was measured using charged
tracks from many consecutive events, 
thus following any significant shifts in beam position
during a LEP fill~\cite{beamspot}. 

All good tracks in one hemisphere were fitted to a common vertex in 
three dimensions. The tracks were required 
to satisfy various requirements to ensure they were well measured, 
including having at least 20
hits in the jet chamber, a transverse momentum with respect to the
beam axis of at least $0.15\rm\,GeV$ and a distance of closest approach to the
beam spot in the $r$-$\phi$ plane of less than 5\,cm. Tracks with a large 
$\chi^2$ contribution to the primary vertex 
fit were removed one by one until each
remaining track contributed less than 4 to the $\chi^2$. In about
$0.7\rm\,\%$ of the events, no tracks remained after this procedure, in
which case the beam spot position was used.

Using the hemisphere primary vertex reconstruction, the primary vertex
resolution in Monte Carlo simulated events with the 1994 silicon
detector is about $70\,\rm\mu m$ in $x$, $15\,\rm\mu m$ in $y$
(where it is dominated by the beam spot constraint) and 
$110\rm\,\mu m$ in $z$. Using the event primary vertex reconstruction,
the resolutions in $x$, $y$ and $z$ are $50\,\rm\mu m$, $15\,\rm\mu m$
and $80\,\rm\mu m$  respectively. The poorer primary vertex resolution
in the hemisphere reconstruction method leads to a non-b impurity
that is about $10\,\%$ higher (relative) for a given b-tagging
efficiency, but a much reduced hemisphere efficiency correlation and associated
systematic error.

\subsection{Secondary Vertex Reconstruction}\label{ss:svtx}

A three dimensional version of the algorithm described in \cite{pr188}
was used to reconstruct a secondary vertex in each jet.
All tracks in the jet with momentum $p>0.5\rm\,GeV$, a distance of closest
approach to the hemisphere primary vertex in the $r$-$\phi$ plane 
(`impact parameter') $d_0<0.3\rm\,cm$, and an impact parameter error
$\sigma_{d_0}<0.1\rm\,cm$, were fitted to a common vertex in three
dimensions. These requirements preferentially select 
vertices from b hadron decays 
because of the high decay multiplicity and hard fragmentation
of the b quark.
Tracks with a large $\chi^2$ contribution to the fit were
removed one by one  until each track contributed less than 4 to the
overall $\chi^2$. At least three tracks were required to remain in the
fit for the secondary vertex finding to be considered successful.

The vertex decay length $L$ was calculated as the length of the vector
from the primary to the secondary vertex, the vector being 
constrained to lie along the
direction of the jet axis. $L$ was given a positive sign 
if the secondary vertex was displaced from the primary in the 
direction of the jet momentum, and a negative sign otherwise. 
Vertices with $L>0$ (`forward') were used in the tagging of b hadron 
decays, and those
with $L<0$ (`backward') were used to reduce the systematics associated with
the detector resolution.

\subsection{Vertex Tagging Neural Network}\label{ss:vmass}

Jets containing a vertex with at least three tracks and satisfying 
$|L/\sigma_L|>3$ (where $\sigma_L$ is the estimated error on $L$,
derived from the track error matrices)  were considered pre-selected 
and input to a neural network algorithm \cite{jetnet}. In the Monte Carlo, the 
efficiency of this pre-selection for b-jets is about 49\% and the 
b-purity is about 76\%. 

The neural network was used to enhance the
b-purity, and has five input nodes, eight hidden nodes and one output
node. The first three inputs were derived 
directly from the reconstructed
secondary vertex: the decay length significance $L/\sigma_L$, the decay
length $L$, and the number of tracks in the secondary vertex $N_s$.
The fourth input to the network tests the stability of the vertex
against mis-measured tracks. The secondary vertex track with the highest impact
parameter significance with respect to the primary 
vertex ($d_0/\sigma_{d_0}$) was
removed from the secondary vertex, and the secondary vertex fit repeated
resulting in the `reduced' decay length significance
\lsred. For genuine b hadron decays, \lsred\ is
large, whilst for vertices caused by one high impact parameter
mis-measured track (which is removed from the `reduced' vertex), \lsred\
is small. 

The fifth input exploits the high mass of b
hadrons compared with charm hadrons, using a method similar to that
described in \cite{alephonetag}. For each track in the jet, a weight $X$
that it came from a b hadron decay was computed, using a separate
artificial neural network 
trained on \bbbar\ Monte Carlo events. This network has
six inputs, eight hidden nodes and one output node. The inputs are:
the scaled track momentum $x_p=p/E_{\rm beam}$; the sine of the angle of
the track with respect to the jet axis
$\sin\theta_t=p_t/p$; the impact parameter significances 
of the track with respect to the reconstructed secondary vertex in the
$r$-$\phi$ ($(d_0/\sigma_{d_0})_{\rm sec}$) and $r$-$z$ 
($(z_0/\sigma_{z_0})_{\rm sec}$) planes; and
the impact parameter significances of the track with respect to
the hemisphere primary vertex in the 
$r$-$\phi$ ($(d_0/\sigma_{d_0})_{\rm prim}$) and $r$-$z$ 
($(z_0/\sigma_{z_0})_{\rm prim}$) planes. The network was trained separately
on tracks with no silicon hits, silicon $r$-$\phi$ hits only, and both
silicon $r$-$\phi$ and $r$-$z$ hits. The output $X$ of this network peaks close
to zero for tracks from fragmentation, and close to one for tracks
from b hadron decay.

All tracks within a jet were then ordered by decreasing $X$ 
({\em i.e.\/} most b hadron
decay-like tracks first), and the first two were clustered together.
Other tracks were added in turn to the cluster, until
the invariant mass of the tracks (assuming them to be pions)
exceeded the charm hadron mass, taken to be $m_D=1.9\rm\,GeV$.
The value $X_D$ of $X$ for the track which caused the cluster
invariant mass to exceed $m_D$ was then used as the fifth input for the
vertex neural network. For charm hadron decays, $X_D$ is
usually small, since tracks from fragmentation  have to be
included to exceed $m_D$. In contrast, b hadrons have enough mass that the 
threshold can usually be exceeded with tracks from the b hadron decay
alone, leading to a value of $X_D$ close to one. If the cluster mass
did not exceed $m_D$ after all tracks in the jet had been added,
$X_D$ was set to zero; this happens more often in charm and light
quark jets than in b jets.

\subsection{Vertex Tag Definition}

Samples of five-flavour Monte Carlo jets passing the pre-selection
and with $L>0$
were used to train the main neural network (with inputs
$L/\sigma_L$, $L$, $N_s$, \lsred\ and $X_D$), to produce output distributions
peaking close to zero for light flavour and charm jets, and outputs
close to one for b jets. 

In order to reduce sensitivity to the modelling of the detector
resolution, the technique of `folding', {\em i.e.\/}
subtracting the number of hemispheres tagged with negative $L$, was employed
\cite{pr188}. This technique works well if the tagging variable (in 
this case the vertex tag neural network output) is  symmetric
about zero for jets containing no particles with detectable lifetime.
To achieve this symmetry, the variables $L/\sigma_L$, $L$ and \lsred\ 
were modified before being input to the neural network. For
$L/\sigma_L$ and $L$ , the absolute values $|L/\sigma_L|$ and 
$|L|$ were taken, while \lsred\ was signed positive if it originally had the
same sign as $L$, and negative otherwise. The impact parameter
significances used to calculate $X_D$ also had their signs reversed 
if $L$ was negative. The magnitude of the vertex tagging variable $B$
was then defined as $|B|=-\ln (1-b)$ where $b$ is the raw neural
network output (between zero and one), and the sign of $B$ was taken to be the
sign of $L$. The logarithmic transformation is used to expand the scale
of the tagging variable in the region just below $b=1$.

The use of folding also requires that
equations~\ref{e:ntsimp} and~\ref{e:nttsimp} are modified
appropriately. The tagging efficiencies $\epsilon$ are replaced by the
difference of forward and backward tagging efficiencies 
$\epsilon_{\rm v}-\epsilon_{\rm\overline{v}}$. The number of
tagged hemispheres \nt\ is replaced by the difference between 
the numbers of forward and backward tagged hemispheres $\nv-\nb$, and
the number of double tagged events \ntt\ is replaced by
$\nvv-\nvb+\nbb$ where \nvv , \nvb\ and \nbb\ are the numbers of events
with two forward tags, one forward and one backward tag, and two
backward tags  respectively \cite{pr188}.

As the silicon detector did not provide $r$-$z$ information in 1992, a
separate version of the vertex tagging algorithm was used for these data.
Primary and  secondary vertices were reconstructed in the  $r$-$\phi$
plane only, and the $r$-$z$ impact parameter significances were not
used in the calculation of  $X_D$. 
According to the Monte Carlo, the tagging impurity for 1992 data using
the $r$-$\phi$ only tagging algorithm is about 30\,\% higher than that
for 1994 data at the same b-tagging efficiency.

\subsection{Vertex Tag Performance}

The distributions of the input variables for the track neural network
used to derive $X_D$ are shown in Figure~\ref{f:trtin94}, 
and the distributions of the vertex tag neural network input variables are
shown in Figure~\ref{f:vnnin94}, for 1994 data and Monte Carlo
simulation. In general the modelling of the input variables is good,
but some discrepancies are visible, particularly 
in the distributions of $N_s$ and
$X_D$. These are attributed to imperfections in the detector
modelling and in the simulation of b hadron decays. The effects of the
former are addressed in the systematic errors, whilst the latter have
no impact on the result since the b-tagging efficiency is determined from
the data and not from the Monte Carlo. 

\epostfig{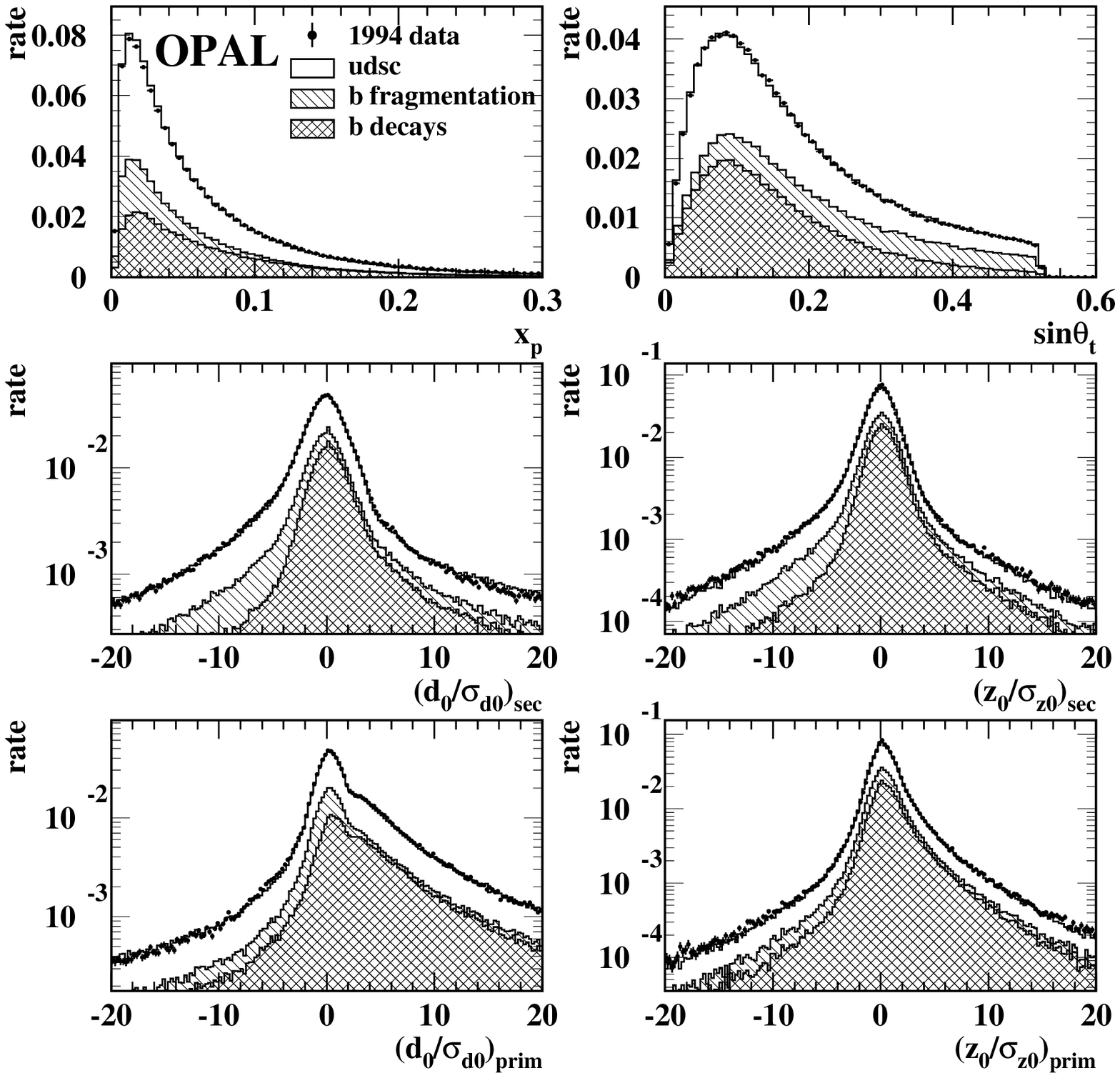}{f:trtin94}{Distributions of the track neural
  network input variables
  $x_p$, $\sin\theta_t$, $(d_0/\sigma_{d_0})_{\rm sec}$,
  $(z_0/\sigma_{z_0})_{\rm sec}$, $(d_0/\sigma_{d_0})_{\rm prim}$
  and $(z_0/\sigma_{z_0})_{\rm prim}$ used to derive $X_D$.
  The sum of the distributions
  for the three track classes (no silicon hits, silicon $r$-$\phi$
  hits only, and both silicon $r$-$\phi$ and $r$-$z$ hits) are shown
  for 1994 data (points) and 1994 Monte Carlo simulation (histograms).
  The estimated contribution of tracks from b decay, b fragmentation,
  and charm and light quark background are indicated.}

\epostfig{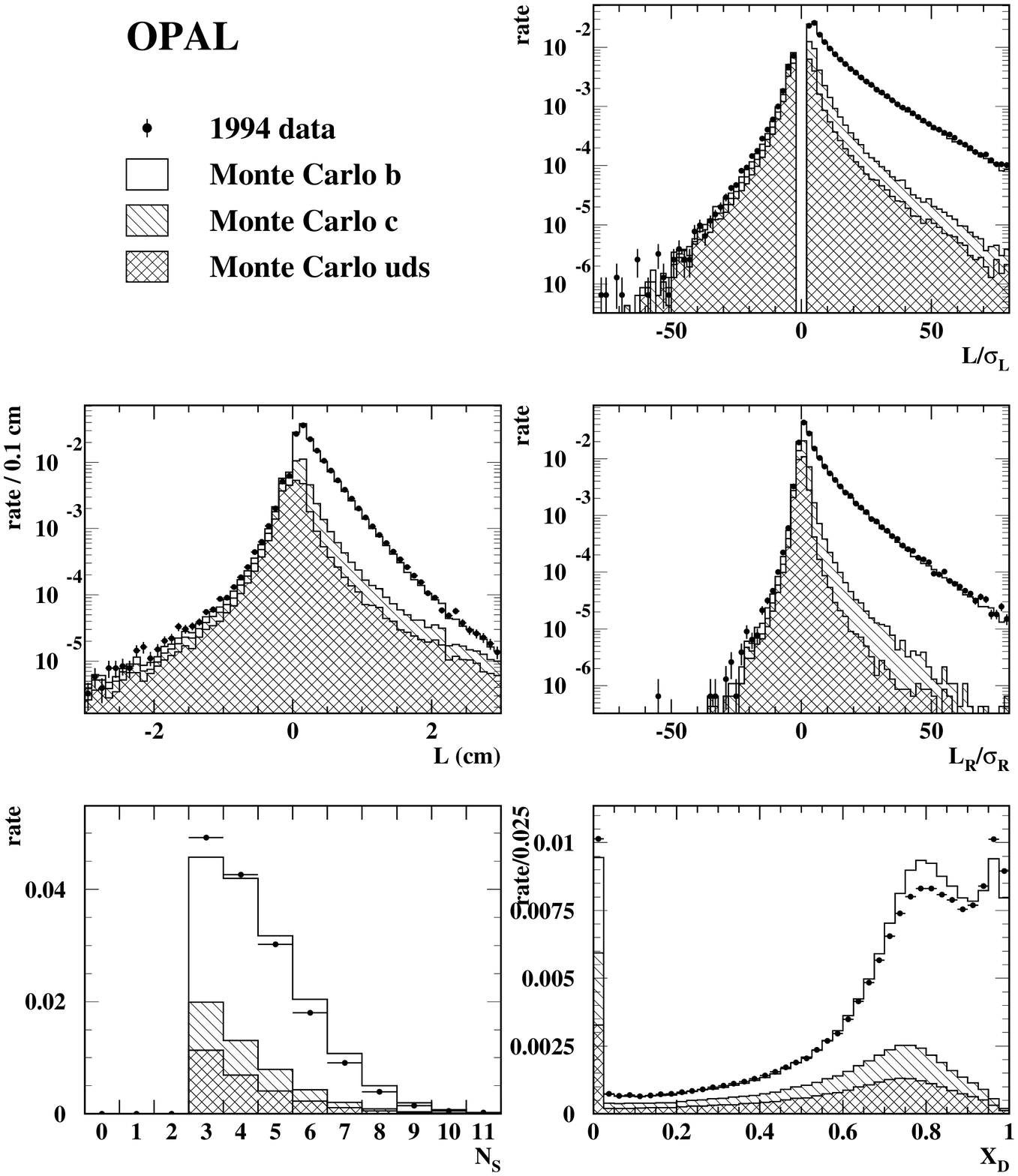}{f:vnnin94}{Distributions of the vertex tagging
  neural network inputs $L/\sigma_L$, $L$, \lsred , $N_s$ and $X_D$
for 1994 data (points) and 1994 Monte Carlo simulation
(histogram). 
The contributions from uds, c and b jets are indicated.}

The distribution of the vertex
tagging variable $B$ is shown for 1994 data and Monte Carlo
in Figure~\ref{f:vnnout94}. Good qualitative 
agreement is seen. The Monte Carlo 
distribution for uds jets is seen to be approximately symmetric about $B=0$, 
as required for the folding procedure to work well. The asymmetry in
the uds distribution is caused by jets containing a long-lived
strange particle ($\rm K^0_S$, $\Lambda$ or other hyperon) and by uds jets
containing a gluon splitting to a \bbbar\ or \ccbar\ pair. 
The distributions for
charm and especially for b jets are shifted to large positive
values of $B$, and at high values the b jet purity exceeds 99\,\%.

\epostfig{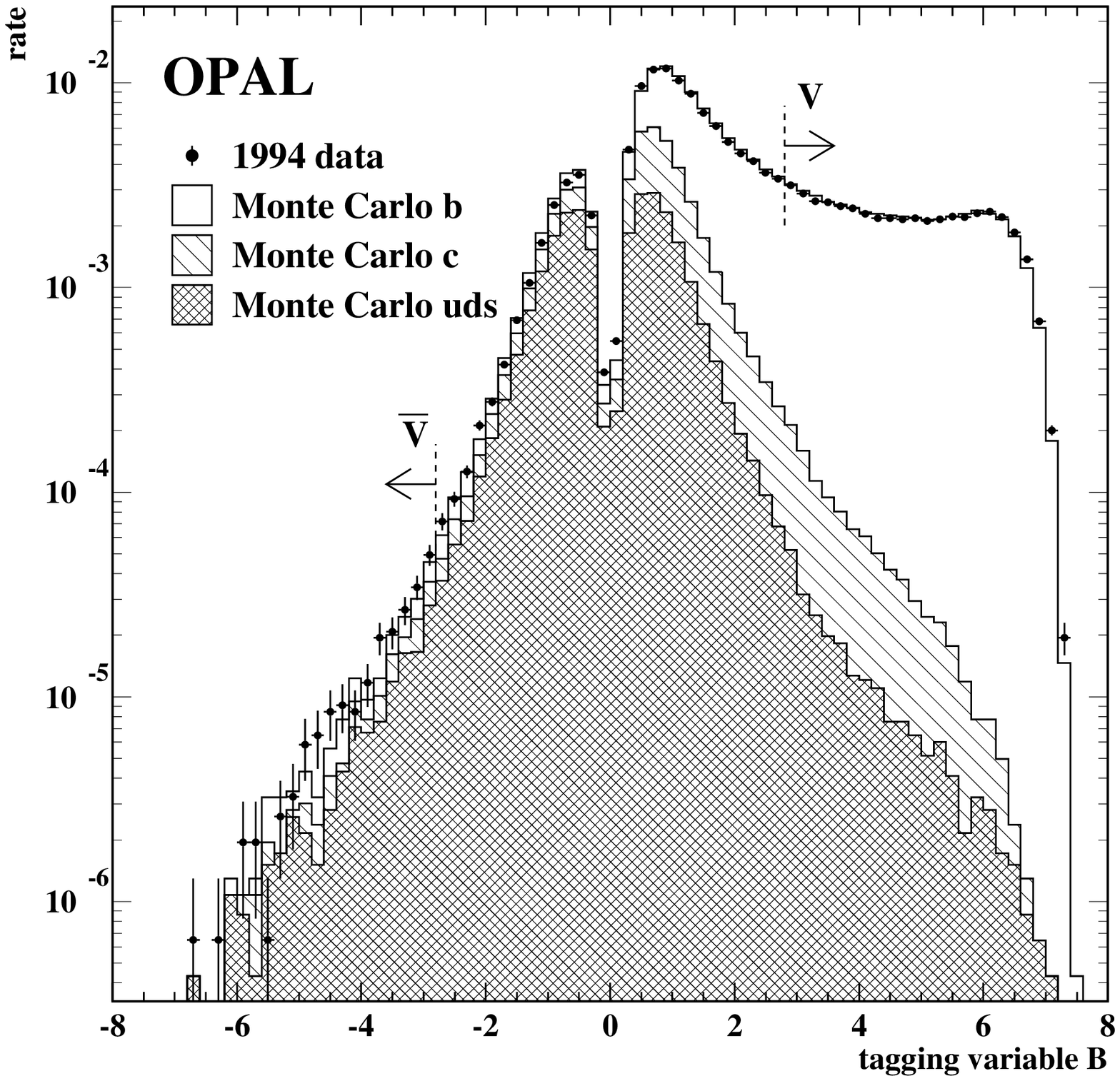}{f:vnnout94}{Distributions of the
  vertex tagging variable $B$ for the 1994 data (points)
  and 1994 Monte Carlo simulation (histogram). 
The positions of the forward and backward tag cuts are shown by dashed lines.
The contributions from uds, c and b jets are indicated.}

The hemisphere vertex tag was defined from the vertex tag of any jet in
the hemisphere. If more than one jet in the hemisphere passed the
vertex tag preselection, the one with the highest value of $|B|$ was
used. A hemisphere was then defined to be 
forward tagged with the vertex tag V if 
$B>\cutv$ and backward tagged (\vbar ) if $B<-\cutv$. The value of 
$\cutv$ was chosen to minimise the overall error on \rb\ when
also including the lepton tag.

\section{Lepton Tagging}\label{s:lepton}

Leptons with high momentum $p$, and a large momentum component
transverse to the jet axis $p_t$, are expected to come mainly from 
semileptonic decays of b hadrons, because of the hard fragmentation
and high mass of the b quark. Electron candidates with $p>2\rm\,GeV$ and
$p_t>1.1\rm\,GeV$, and muon candidates with $p>3\rm\,GeV$ 
and $p_t>1.4\rm\,GeV$,
were used to tag \bbbar\ events. Both electrons and muons were
restricted to the polar angle range $|\cos\theta|<0.8$ to ensure a
well-defined acceptance and good Monte Carlo modelling of the
efficiencies and backgrounds.

The sources of lepton candidates are
divided into two classes. The first class consists of prompt
leptons from the decays of b and c hadrons (including $\rm
b\rightarrow\tau\rightarrow\ell$ and $b\rightarrow
J/\psi\rightarrow\ell$ and those from b and c hadrons produced in
gluon splitting). These leptons are included in the definitions
of \eb, \ec\ and \euds\ for the lepton tag. Monte Carlo is used to
estimate \ec\ and \euds, whereas \eb\ is determined from the data.
The second class of lepton candidates consists of everything
else: real leptons produced from Dalitz decays, 
photon conversions and the decays in
flight of $\rm K^\pm$ and $\pi^\pm$, and hadrons mis-identified as leptons.
This  background is estimated using a combination of data and
Monte Carlo, and is subtracted from the number of identified lepton candidates
before input to the fit for \rb. These lepton candidates  are therefore not
included in the definitions of \eb, \ec\ and \euds.

\subsection{Electron Identification}\label{ss:elid}

Electrons were identified using an artificial neural network.
The algorithm is a simplified version of that described in 
\cite{elecid}, using only six rather than twelve neural network
inputs. The inputs are: the momentum and polar angle of the track, the
energy-momentum ratio $E/p$, the number of electromagnetic calorimeter
blocks contributing to the energy measurement, the normalised
ionisation energy loss 
\dxnorm\ and the error on the ionisation energy loss $\sigma_{\dedx}$. The
normalised $\dedx$ value is defined as
$\dxnorm=(\dedx-(\dedx)_0)/(\sigma_{\dedx})_0$, where $(\dedx)_0$ is the
value and $(\sigma_{\dedx})_0$ the error expected for the track,
assuming it to be an electron of the measured track momentum.
The neural network output for
electron candidates in data and Monte Carlo is shown in 
Figure~\ref{f:elnn}(a). Electron candidates were required to
have a neural network output greater than 0.95.

\epostfig{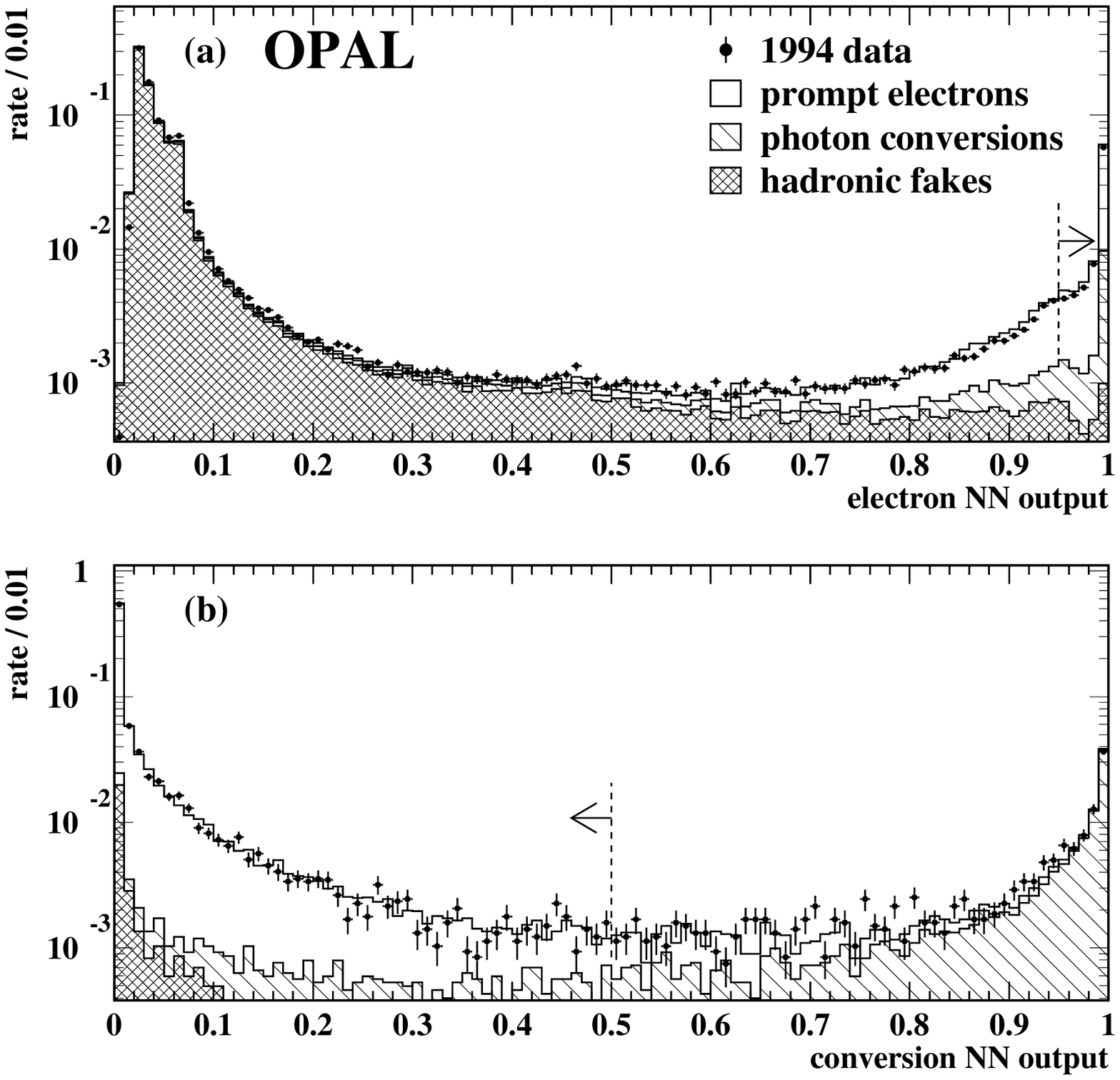}{f:elnn}{Performance of neural network electron
  identification in data and Monte Carlo: (a) Normalised distributions of
  electron neural network output $N_{\rm el}$ 
  for all tracks with $\dxnorm>-2$, $p>2\rm\,GeV$ and $p_t>1.2\rm\,GeV$; 
  (b) Normalised distributions of the photon
  conversion tagging neural network output for identified electron
  tracks with $N_{\rm el}>0.95$. In each case, the 1994 data is shown
  by the points with error bars, and the Monte Carlo contributions
  from prompt electrons, photon conversions and mis-identified hadrons 
  by the histograms. The selected regions are shown by the dashed lines
  and arrows.}

Photon conversion candidates were rejected using another
neural network algorithm, using  spatial
matching, invariant mass and momentum information of the electron
candidate and an oppositely charged partner track. This algorithm is
similar to that described in \cite{elecid}, but uses the new electron neural
network algorithm described above. The distributions of this neural network
output are shown in Figure~\ref{f:elnn}(b).
Electron candidates were required to have a
conversion tagging neural network output smaller than 0.5.
After all these requirements, the expected
identification efficiency for electrons from decays of b hadrons
within the kinematic and geometrical acceptance is about $68\,\%$.

The non-prompt background in the tagged electron sample consists of
hadrons mis-identified as electrons, untagged photon conversions and a
small number of electrons from Dalitz decays of light mesons.
The background mis-identification probabilities were found to depend strongly
on the $p$ and $p_t$ of the track, but were otherwise largely independent of
the event flavour.  Since the track $p$ and $p_t$
distributions are different in bottom, charm and light quark events,
the number of fake leptons in hemispheres opposite tagged and untagged
hemispheres are different. Therefore, the Monte Carlo was used to
determine the fake probability per track as a function of $p$ and
$p_t$. The probabilities were combined with the two
dimensional distributions of track $p$ and $p_t$ measured 
opposite untagged and tagged hemispheres in the data, to estimate the
number of fake electrons expected opposite each type of hemisphere.
These estimates were then subtracted from the number of lepton tagged 
hemispheres and events before input to the fit for \rb.

In total, $26185$ hemispheres were tagged by electrons after the photon
conversion rejection. Of these, $918 \pm 30$ were attributed to hadronic
fakes, $784 \pm 28$ to untagged photon conversions and $210\pm 14$ to 
Dalitz decays of light mesons, where the background rates have
been evaluated by subdividing the fake probabilities by source as a
function of $p$ and $p_t$. The errors are due to data statistics only.

\subsection{Muon Identification}\label{ss:muid}

Muon candidates were identified by matching track segments reconstructed in
the four-layer external muon chambers to tracks extrapolated from the
central tracking detectors. The measured $\dedx$ was also required
to be consistent with a muon. The algorithm is described in detail 
in \cite{muonid}. The expected identification efficiency
for muons from decays of b hadrons within the kinematic and
geometrical acceptance is about $79\,\%$.

The muon background was estimated from the Monte Carlo using the same
techniques as  described in Section~\ref{ss:elid}. 
In total, $21558$ hemispheres containing muon candidates were found in
the data, of which $3311\pm 58$ were attributed to hadronic background. The
errors are due to data statistics only.

\section{Measurement of \rb } \label{s:meas}

The numbers of hadronic events, 
tagged hemispheres and double tagged events found in 
each year of the data are listed in Table~\ref{t:evct}. 
The symbol $N_i$ represents the number of hemispheres
tagged by tag $i$, where $i=\rm v$ for
forward tagged vertices, $i=\rm\overline{v}$ for backward vertices, 
$i=\ell$ for leptons and $i=a$ for hemispheres tagged by either a
forward vertex or a lepton. The symbol $N_{ij}$ represents 
the number of events tagged by tag $i$ in one
hemisphere and tag $j$ in the other hemisphere. The expected photon
conversion and hadronic backgrounds have already been subtracted from
the lepton counts as described in Section~\ref{s:lepton}.

\begin{table}
\centering 

\begin{tabular}{ll|rrrr}\hline \hline
& & 1992 & 1993 & 1994 & 1995 \\
\hline
Number of events & $N_{\rm had}$ & 373462 & 401674 & 770366 & 377738 \\
\hline
Tagged hemispheres 
& $\nv-\nb$ & 31455 & 35942 & 71624 & 36248 \\
& $\nl$     & 7726 & 8798 & 16901 & 8498 \\ 
& $\na-\nb$ & 37877 & 43086 & 85145 & 43069 \\
\hline
Double tagged events 
& $\nvv-\nvb+\nbb$ & 2934 & 3562 & 7376 & 3964 \\
& $\nll$           & 161 & 171 & 379 & 191 \\
& $\nvl-\nbl$      & 1123 & 1329 & 2496 & 1289 \\
& $\naa-\nab+\nbb$ & 4162 & 4993 & 10101 & 5369 \\
\hline
\end{tabular}
\caption{\label{t:evct}
Numbers of hadronic events, tagged hemispheres and double tagged
events selected in each year of the data. Background has been
subtracted from the lepton samples, and the resulting counts are
quoted to the nearest whole number.}
\end{table}

The hemisphere tagging probabilities for charm and 
light quark events, estimated using Monte Carlo simulation, are given in
Table~\ref{t:udscefi}. The tag probabilities vary slightly from year to
year because of the differences in silicon geometrical acceptance
and the $r$-$\phi$ only  tag
used in 1992. The Monte Carlo predicted no significant difference in tagging
efficiency between $\rm u\overline{u}$, $\rm d\overline{d}$ and
$\rm s\overline{s}$ events, and the effect on \rb\ of assuming a common
efficiency was estimated to be less than $10^{-6}$.

\begin{table}
\centering

\begin{tabular}{l|c|cc}\hline \hline
Tag & Year & \ccbar\,(\%) & \udsbar\,(\%) \\
\hline
Vertex   & 1992 & $0.463\pm 0.009$ & $0.0279\pm 0.0034$ \\
         & 1993 & $0.457\pm 0.015$ & $0.0256\pm 0.0020$ \\
         & 1994 & $0.473\pm 0.007$ & $0.0256\pm 0.0014$ \\
         & 1995 & $0.491\pm 0.009$ & $0.0253\pm 0.0023$ \\
\hline
Lepton   & 1992 & $0.425\pm 0.009$ & $0.0181\pm 0.0016$ \\
         & 1993 & $0.455\pm 0.013$ & $0.0196\pm 0.0010$  \\
         & 1994 & $0.443\pm 0.006$ & $0.0172\pm 0.0006$ \\
         & 1995 & $0.445\pm 0.008$ & $0.0155\pm 0.0010$ \\
\hline
Combined & 1992 & $0.890\pm 0.013$ & $0.0467\pm 0.0037$ \\
         & 1993 & $0.902\pm 0.016$ & $0.0425\pm 0.0030$ \\
         & 1994 & $0.918\pm 0.009$ & $0.0425\pm 0.0015$ \\
         & 1995 & $0.936\pm 0.012$ & $0.0405\pm 0.0025$ \\
\hline
\end{tabular}
\caption{\label{t:udscefi}
Hemisphere tagging probabilities for charm and light quark events
  estimated from the Monte Carlo for each year of the data. The errors
  are due to Monte Carlo statistics only.}
\end{table}

The \bbbar\ tagging efficiency correlations $\cb-1$ for each year of
the data are listed in Table~\ref{t:corl1}. These values were
determined from a Monte Carlo sample of 7~million \bbbar\ events, by
calculating the ratio of the event double tagging probability to the product
of the hemisphere single tagging probabilities. Small corrections
for the differing detector performance in each year were derived from
the data, as will be discussed in Section~\ref{ss:geoc}. The errors include
contributions from the Monte Carlo and data statistics only.

\begin{table}
\centering

\begin{tabular}{l|ccc}\hline\hline
$\cb-1$ (\%) & Vertex & Lepton & Combined \\ \hline
1992 & $0.84\pm 0.24$ & $1.82\pm 1.29$ & $1.01\pm 0.21$ \\
1993 & $1.10\pm 0.26$ & $1.34\pm 1.28$ & $1.13\pm 0.20$ \\
1994 & $0.77\pm 0.22$ & $1.13\pm 1.25$ & $0.92\pm 0.18$ \\
1995 & $0.99\pm 0.24$ & $0.92\pm 1.28$ & $0.95\pm 0.19$ \\
\hline
\end{tabular}
\caption{\label{t:corl1} Hemisphere efficiency correlations in
  \bbbar\ events estimated from the Monte Carlo with small corrections
  for each year of the data. The errors are due to Monte Carlo and
  data statistics only.}
\end{table}

The values of \rb\ and the tagging efficiencies \eb\ 
derived for each year of the data are
given in Table~\ref{t:rbefdata}. The results are obtained by solving
equations~\ref{e:ntsimp} and~\ref{e:nttsimp}.
The full result from the combined tag
is calculated using the number of tagged hemispheres $\na-\nb$ and the
number of double tagged events $\naa-\nab+\nbb$. Results are also
given for the vertex tag alone ($\nv-\nb$ and $\nvv-\nvb+\nbb$) and
the lepton tag alone ($\nl$ and $\nll$). The results  have been 
corrected for the event selection bias described in Section~\ref{s:dsam}.
Only the data statistical errors are included.
The results from the combined tag for the four years agree with each
other at a $\chi^2$ of $1.8$ for 3~degrees of freedom;
the corresponding values of $\chi^2$ for the vertex
and lepton tags alone are $3.4$ and $2.9$ respectively.
Combining the data from all four years, the value of \rb\ is measured to be:
\[
\rb = \rbval\pm \rbstat
\]
where the error is due to the data statistics only. 

\begin{table}
\centering

\begin{tabular}{lr|cc|c}\hline\hline
 &  & Vertex & Lepton & Combined \\
\hline
\rb
& 1992 & $0.2169\pm 0.0033$ & $0.2156\pm 0.0174$ & $0.2164\pm 0.0027$ \\
& 1993 & $0.2182\pm 0.0030$ & $0.2495\pm 0.0199$ & $0.2182\pm 0.0025$ \\
& 1994 & $0.2176\pm 0.0021$ & $0.2113\pm 0.0110$ & $0.2194\pm 0.0018$ \\
& 1995 & $0.2120\pm 0.0027$ & $0.2171\pm 0.0160$ & $0.2157\pm 0.0023$ \\
& Combined & $0.2163\pm 0.0013$ & $0.2184\pm 0.0074$ & $0.2178\pm 0.0011$ \\
\hline
\eb
& 1992 & $0.1893\pm 0.0030$ & $0.0439\pm 0.0035$ & $0.2254\pm 0.0029$ \\
& 1993 & $0.2002\pm 0.0029$ & $0.0406\pm 0.0032$ & $0.2369\pm 0.0028$ \\
& 1994 & $0.2087\pm 0.0021$ & $0.0477\pm 0.0025$ & $0.2429\pm 0.0020$ \\
& 1995 & $0.2211\pm 0.0030$ & $0.0477\pm 0.0035$ & $0.2550\pm 0.0028$ \\
\hline
\end{tabular}
\caption{\label{t:rbefdata}
Values of \rb\ and \eb\ measured in each year of the data, 
after correlation and event selection  bias correction. Only
statistical errors are included.}
\end{table}

The result depends on \rc\ as follows:
\[
\frac{\Delta\rb}{\rb}=\delrbrc \frac{\Delta\rc}{\rc}
\]
where $\Delta\rc$ is the deviation of \rc\ from the value 0.172
predicted by the Standard Model and used in this analysis.

The systematic errors coming from sources other than \rc\ are discussed
below and are summarised in Table~\ref{t:systsum}. Most of the
systematic errors arise through the charm and light quark hemisphere tagging
efficiencies, \ec\ and \euds, and through the \bbbar\ tagging
efficiency correlation \cb. The dependence of \rb\ on these quantities
is given approximately by:
\begin{equation}
\frac{\Delta\rb}{\rb}=\delrbec\frac{\Delta\ec}{\ec}
\delrbeuds\frac{\Delta\euds}{\euds}+\frac{\Delta\cb}{\cb}\label{e:rbsens}
\end{equation}
The systematic errors arising from the charm and light quark 
efficiencies are discussed in
detail in Section~\ref{s:syste}, and those from the efficiency
correlation in Section~\ref{s:systc}.
The systematic errors on the efficiencies are also given in 
Table~\ref{t:systsum}. The only other source of systematic error is
the hadronic event selection, which was discussed in
Section~\ref{s:dsam} and gives rise to an error of $\pm\sysevtsel$ on \rb.

\begin{table}[t]
\centering

\begin{tabular}{l|cc|c}\hline \hline
Source & $\Delta\ec /\ec$ (\%) & $\Delta\euds /\euds$ (\%) &
$\Delta\rb$ \\ 
\hline
Tracking resolution & 1.24 & 4.0 & \systrkres \\
Tracking efficiency & 0.80 & 4.0 & \systrkefi \\
Silicon hit matching efficiency & 0.82 & 2.8 & \syssidrop \\
Silicon alignment & 0.58 & 2.1 & \syssialgn \\
Electron identification efficiency & 1.11 & 0.5 & \syselidef \\ 
Muon identification efficiency & 0.64 & 0.2 & \sysmuidef \\ 
c quark fragmentation & 2.26 & - &  \syscqfrag \\
c hadron production fractions & 3.66 & - & \syschprod \\
c hadron lifetimes & 0.55 & - & \syschlife \\
c charged decay multiplicity & 1.09 & - & \syschmult \\
c neutral decay multiplicity & 2.39 & - & \syschneut \\
Branching fraction $B(\rm D\rightarrow K^0)$ & 1.20 & - & \syschklam \\
c semileptonic branching fraction & 2.44 & - & \syscsemil \\
c semileptonic decay modelling & 2.34 & - & \syscsdmod \\
Gluon splitting to \ccbar\  & 0.34 &  6.3 & \sysgsplcc \\
Gluon splitting to \bbbar\  & 0.50 &  9.3 & \sysgsplbb \\
$\rm K^0$ and hyperon production & - & 0.3 & \sysklhypr \\
Monte Carlo statistics (c, uds) & 0.66 & 2.5 & \sysmcstat \\
\hline
Subtotal $\Delta\ec$ and $\Delta\euds$ & 
 6.65 & 13.3 & \sysefitot \\
\hline
\multicolumn{3}{l|}{Electron identification background}  & \syselidbg \\
\multicolumn{3}{l|}{Muon identification background}      & \sysmuidbg \\
\multicolumn{3}{l|}{Efficiency correlation $\Delta\cb$}  & \syscrltot \\
\multicolumn{3}{l|}{Event selection bias}   & \sysevtsel \\
\hline
\multicolumn{3}{l|}{Total} & \rbsysq \\
\hline
\end{tabular}
\caption{\label{t:systsum} Systematic errors on the measured value of
  \rb . The uncertainties on the charm and light quark efficiencies
    for each tag are also given. The systematic errors arising from
    the efficiencies and lepton identification background are
    discussed in Section~\ref{s:syste}, those from efficiency
    correlation in Section~\ref{s:systc} and that from the event selection in
    Section~\ref{s:dsam}.}
\end{table}

As a cross check, the cut on the vertex tag neural network output $B$
was varied in the range 2.2--3.8, and the cuts on the lepton transverse
momenta were varied by up to $\pm 0.3\rm\,GeV$ from their nominal values.
The values of \rb\ obtained are shown
in Figure~\ref{f:rbcut}, together with the uncorrelated parts of the
statistical and systematic errors.  No significant trend in the
measured value of \rb\ is observed within these errors.

\section{Systematic errors: tagging efficiencies}\label{s:syste}

Systematic errors on the tagging efficiencies \ec\ and \euds\ arise
from the understanding of the tracking detectors, the electron and muon
identification, and the various physics parameters input to the Monte
Carlo simulation. 

\subsection{Tracking detector performance}\label{s:detsys}

The evaluation of the charm and light quark tagging efficiencies 
for the vertex tag 
requires an accurate simulation of the detector resolution
for charged tracks. The Monte Carlo simulation has been tuned to
reproduce the tracking resolution seen in each year of the data by
studying the impact parameter distributions of tracks, as functions 
of track momentum, polar angle and the different sub-detectors
contributing hits.

The effect of uncertainties in this procedure was evaluated as follows:
\begin{description}
\item[Tracking resolution:] The sensitivity to the tracking resolution
  was assessed by degrading or improving the resolution 
  in the Monte Carlo. This was done by applying  
  a single multiplicative scale factor $\beta$ to the difference
  between the reconstructed and true track parameters of all charged
  tracks \cite{pr188}. A $\pm 10\,\%$ 
  variation was applied to the $r$-$\phi$ track parameters (the impact
  parameter $d_0$ and track azimuthal angle $\phi_0$). An independent 
  $\pm\,10\,\%$ variation was applied to the analogous parameters in
  the $r$-$z$ plane. Together with uncertainties in the simulation of
  b hadron decays and fragmentation, these variations can account for all the
  discrepancies observed between data and Monte Carlo in the neural network
  input and output distributions. The result of these variations is an
  error of $\pm \systrkres$ on \rb.

\item[Track reconstruction efficiency:] The 
  reconstruction efficiency for charged tracks is estimated to
  exceed 98\,\% for the momentum and impact parameter 
  requirements described in
  sections~\ref{ss:pvtx} and~\ref{ss:svtx}. Most of the losses
  occur in small regions of $\phi$ around the jet chamber cathode and
  anode wire planes. The $\phi$ distribution of reconstructed tracks
  is well reproduced by the Monte Carlo simulation, as is the
  two-track resolution and the rate of tracks lost because they lie on
  top of another track when reflected in the anode plane. Residual
  discrepancies indicate that the track reconstruction efficiency is modelled 
  to within $\pm 1\,\%$. The effect on \rb\ was assessed by randomly
  discarding 1\,\% of tracks in the Monte Carlo, and was found to be
  $\pm \systrkefi$.

\item[Silicon hit association efficiency:] The Monte Carlo was tuned to
  model the overall rate of associating silicon hits to tracks, and to
  model known inefficient regions of the silicon detector in the 1993
  and 1994 data. The residual discrepancies between data and Monte
  Carlo association  efficiencies were found to be within
  $\pm 1\,\%$ for the $r$-$\phi$ and $\pm 3\,\%$ for the $r$-$z$
  hits. The hit association efficiency in the Monte Carlo was 
  varied within these errors resulting in a systematic error on \rb\
  of $\pm \syssidrop$.

\item[Silicon alignment:] The position of the $r$-$\phi$ silicon
  wafers in the azimuthal direction, and the \mbox{$r$-$z$} wafers along the
  $z$ direction, is determined by an
  alignment procedure using $\zb\rightarrow\mu^+\mu^-$ events
  to a precision of about $10\rm\,\mu m$
  \cite{opaldet}. This uncertainty is included in the Monte Carlo
  simulation, but has little impact on the track resolution
  in hadronic events which is dominated by multiple scattering.
  The radial alignment uncertainty is
  much more important since radial shifts of individual wafers can
  lead to systematic mis-measurement of the decay length $L$ in jets
  contained largely within a single wafer. The radial alignment
  was studied using cosmic rays (which are incident on the wafers at
  all angles) recorded throughout the data
  taking period and found to be good to a precision of $\pm 20\rm\,\mu m$.
  The effect on the tagging efficiencies was studied by  
systematically displacing one or both silicon barrels radially by
  $20\rm\,\mu m$ in the simulation, and was found to correspond to an
  error of $\pm \syssialgn$ on \rb.
\end{description}
\epostfig{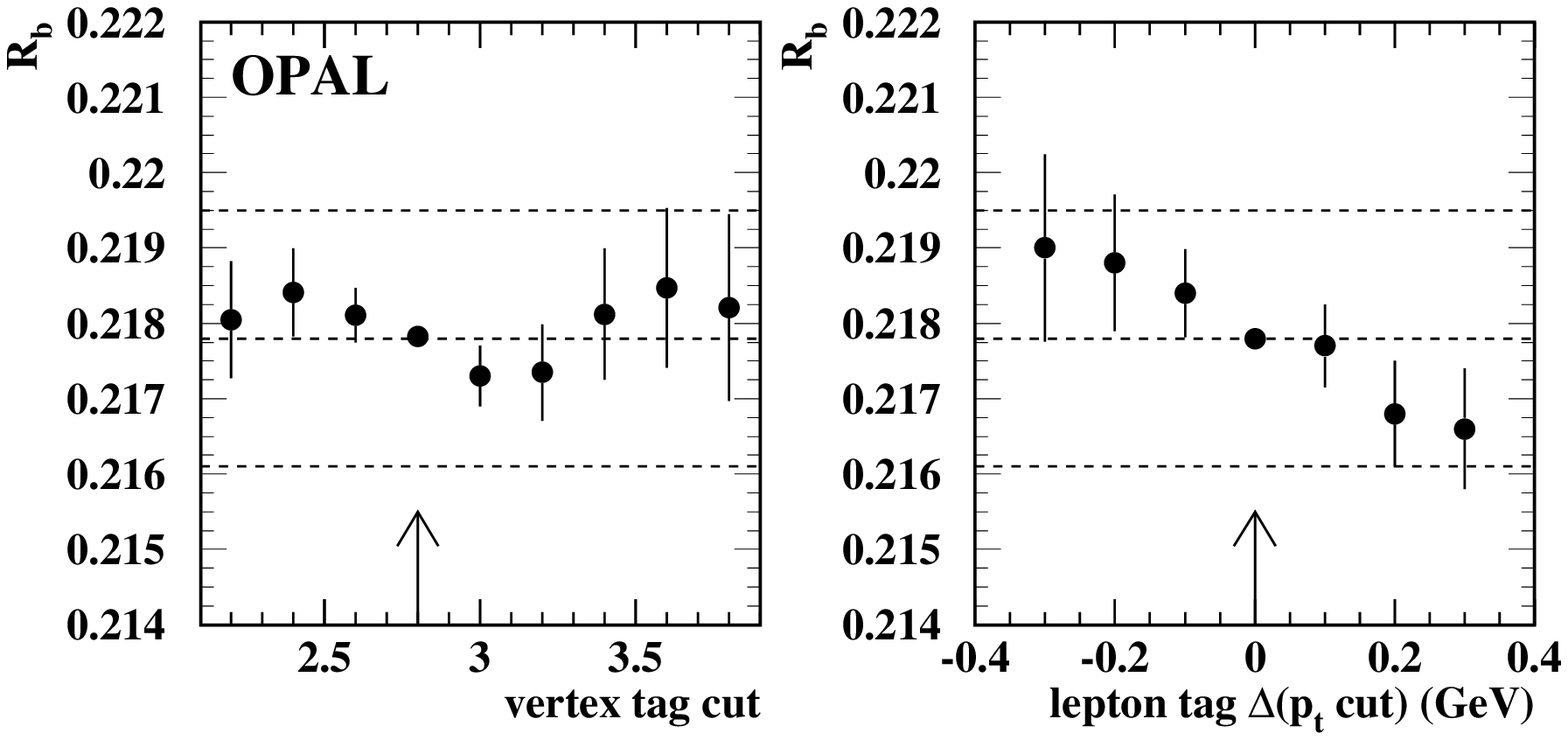}{f:rbcut}{Values of \rb\ obtained at
    different cut values. The central value is indicated by the arrow 
    and the dashed lines indicate the total statistical and systematic
    error. The error bars
    indicate the uncertainties on the differences from the
    central result, including both statistical and systematic components.}

\epostfig{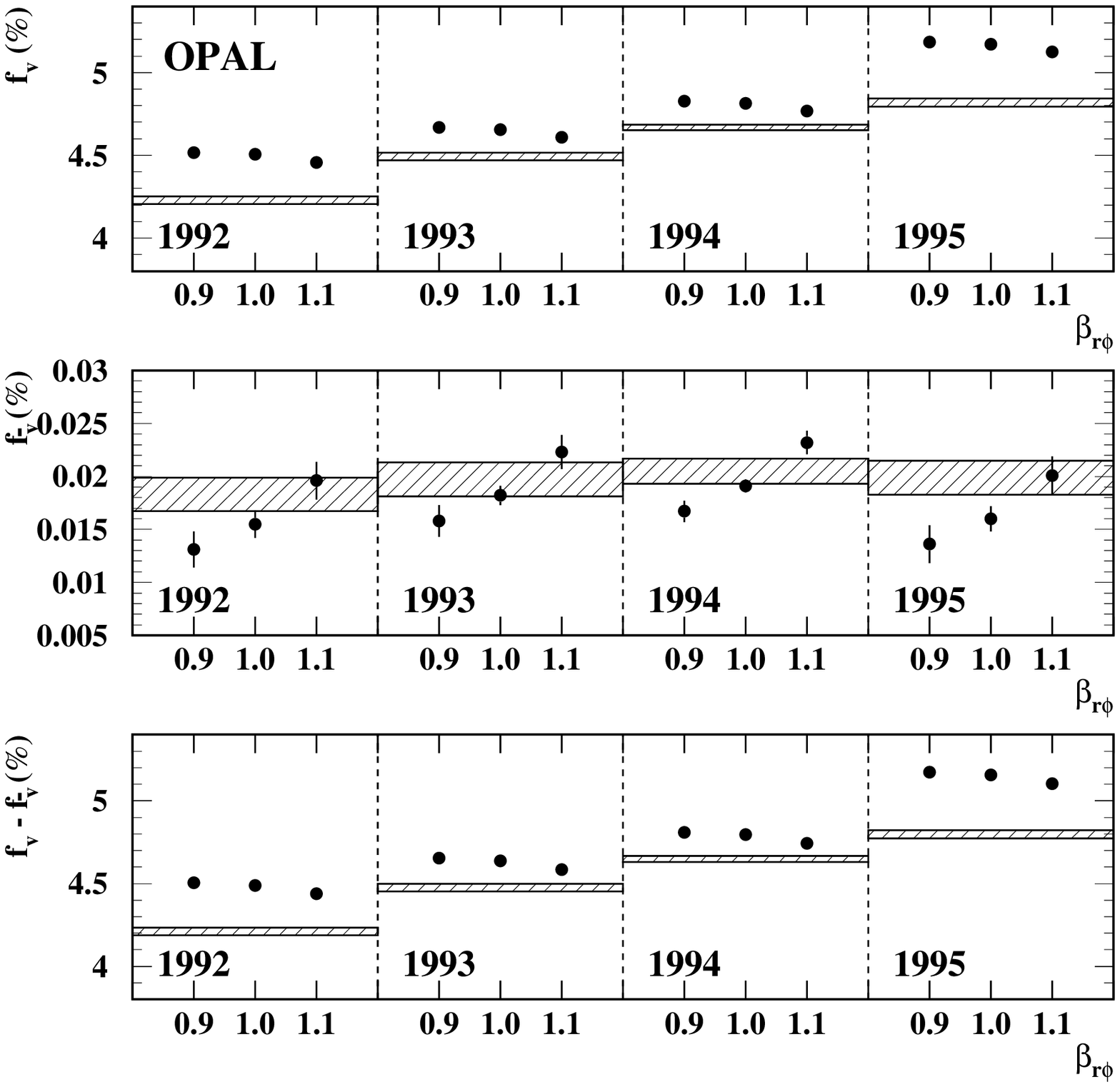}{f:vtxrate}{Forward (\fv ), backward (\fb ) and
  folded ($\fv -\fb$) hemisphere tagging fractions in each year for
  the vertex tag. The data tagging rates are shown by the shaded bars,
  the width of the bars representing the statistical error.
  The Monte Carlo predictions as a function of the
  $r$-$\phi$ resolution scaling parameter $\beta_{r\phi}$  
  are shown by the points with error bars.}

The forward, backward and folded vertex tagging rates measured in each year
of the data are shown in Figure~\ref{f:vtxrate}. 
The Monte Carlo
prediction as a function of the $r$-$\phi$ resolution parameter $\beta$ is
also shown. The Monte Carlo forward tagging rate is not important,
since the b-tagging efficiency is determined directly from the data
using the double tagging technique. It is always somewhat higher in
the Monte Carlo than in the data, but the observed discrepancies are within the
uncertainties associated with b-physics and detector simulation in the
Monte Carlo. The
changes in tagging rates as a function of year are mainly caused by
differences in the silicon detector in each year of data taking. In
1992, only $r$-$\phi$ coordinate information was available, the
$r$-$z$ wafers being installed before the 1993 run \cite{opalsi3d}. A
faulty silicon module in 1993 was replaced for the 1994 run, and a
new detector with increased coverage installed for 1995. 

The Monte Carlo simulation has been tuned by studying the impact
parameter distributions of single tracks. This procedure is not
sensitive to possible coherent effects affecting all the tracks in a 
small number of jets in a correlated way. For example, such  effects can
be caused by a badly mis-reconstructed primary vertex,
which could change the impact parameters of all tracks in a jet, potentially
giving an apparent large negative decay length and hence a backward tag.
However, the discrepancies in the backward tagging rates 
seen in Figure~\ref{f:vtxrate} are 
within the range covered by the $\pm 10\,\%$ variation in the
$r$-$\phi$ resolution parameter $\beta$, so such coherent tracking
resolution effects are not likely to be important. This was also
checked by studying tails in the distributions of differences in the
primary vertex position measured in each hemisphere.

The effect of the $r$-$\phi$ resolution scaling on the light quark and
charm tagging efficiencies is shown in Figure~\ref{f:udscefi}, for
variations in $\beta_{r\phi}$ between 0.8 and 1.2 ($\pm 20\,\%$).
These variations cause large changes in the light quark (uds)
forward and backward tagging efficiencies, but these changes largely
cancel in the folded tagging efficiencies, thus reducing the
systematic uncertainty. In contrast, the charm tagging efficiency is
dominated by real lifetime tags, and the folding procedure has only a
small effect.

\epostfig{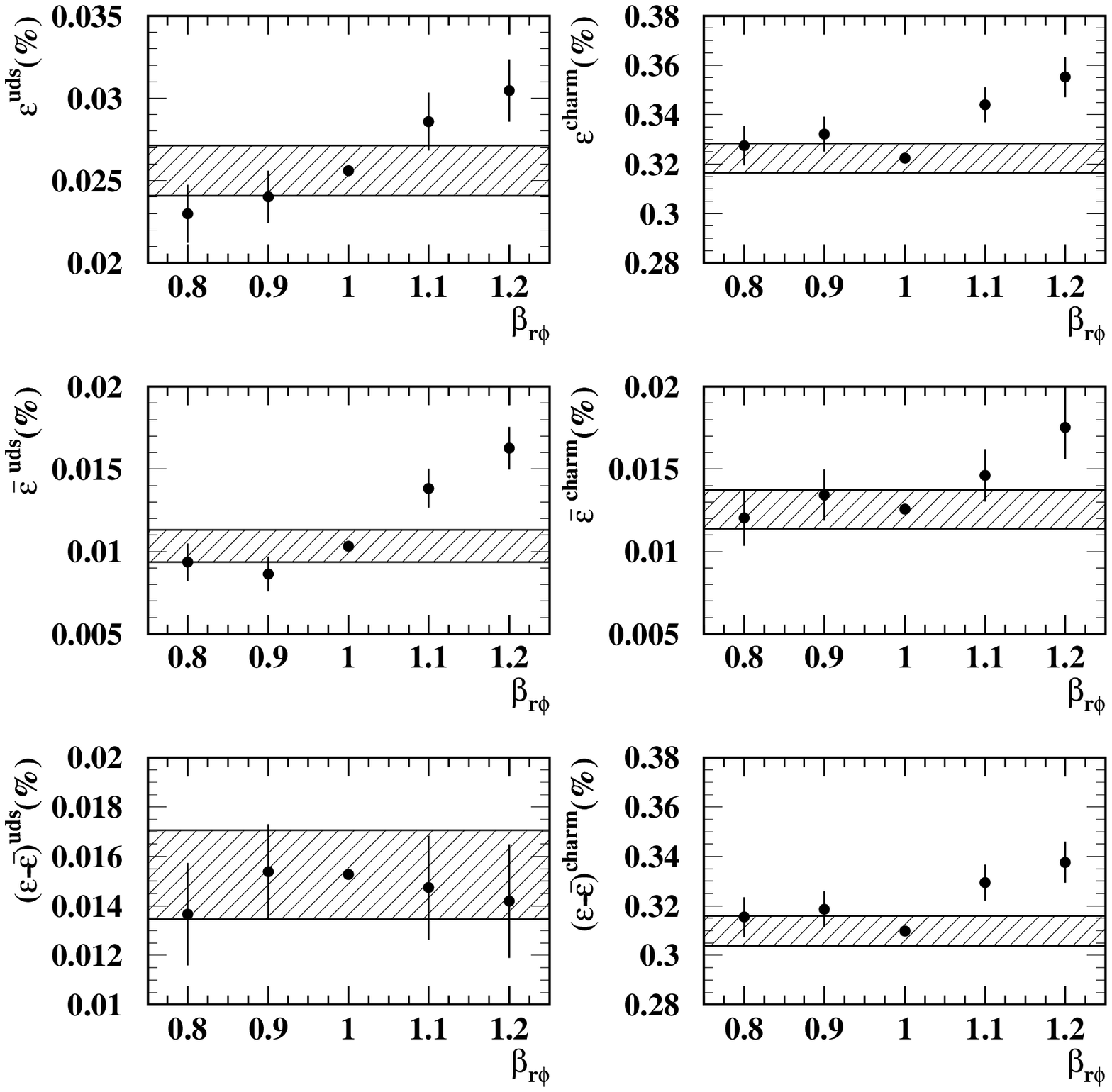}{f:udscefi}{Forward ($\epsilon$), backward 
($\overline\epsilon$) and folded ($\epsilon-\overline{\epsilon}$) vertex
tagging efficiencies, and their differences for light quark (uds) and
charm hemispheres in 1994 Monte Carlo, as a function of the $r$-$\phi$
resolution scaling parameter $\beta_{r\phi}$. 
The shaded bands around the central $\beta_{r\phi}=1.0$ 
points show the total statistical error, and the error bars on the other points
show the statistical errors on the difference between them and the
central points.}

\subsection{Electron identification}

Monte Carlo simulation was used to predict the efficiency for identifying
prompt electrons from $\rm c\rightarrow e$ decays and those from gluon
splitting  $\rm g\rightarrow (\bbbar,\ccbar)\rightarrow e$, and
background rates as function of track $p$ and $p_t$.
The modelling of the electron identification requirements in the Monte
Carlo was checked by studying in detail the distributions of the input
variables of the neural network. The effects on the electron identification
performance of discrepancies due to each input variable were added in
quadrature to estimate the total systematic errors. This procedure
was cross checked with samples of electrons and hadrons selected
from various pure control samples.

The Monte Carlo efficiency prediction was checked by studying the
neural network input distributions and with samples of
pure electrons from $\rm e^+e^-\rightarrow e^+e^-$ events and 
photon conversions, and was found to be modelled
to a relative precision of $\pm \elefierr$. According to the Monte
Carlo, about 2.6\,\% of true prompt
electrons are rejected by the photon conversion finder, and this was
checked to a relative precision of $\pm 30\,\%$ by comparing samples of high
$p$ and $p_t$ electrons in the data and Monte Carlo. This corresponds
to a further error of $\pm 0.8\,\%$ on the prompt electron efficiency.
Combining both the electron identification and photon rejection uncertainties
gives a systematic error on \rb\ of $\pm \syselidef$.

The two main non-prompt backgrounds in the tagged electron sample are
charged hadrons (mainly $\pi^\pm$) which are mis-identified as electrons, and
untagged photon conversions. The effect of mis-modelling of the neural
network input distributions in the Monte Carlo was found to correspond
to an uncertainty of $\pm 21\,\%$ in the hadron mis-identification
probability. Consistent results were found using control samples of
$\pi^\pm$ selected in $\rm K^0_s\rightarrow\pi^+\pi^-$ and $\tau\rightarrow 3\pi$ 
decays. The probability for
mis-identified hadrons to pass the conversion rejection requirements
was found to be modelled to $\pm 5\,\%$ using the 
$\rm K^0_s\rightarrow\pi^+\pi^-$ sample, a sample of prompt muons 
(which have similar $p$ and $p_t$ distributions to prompt electrons
and no photon conversion background) and an
inclusive electron-depleted sample selected using an anti-cut on the 
electron identification neural network.

The untagged photon conversion background was studied using a sample
of identified electrons, enriched in conversions by requiring the
tracks to have low $p$ and $p_t$. The number of untagged conversions
was found to be modelled to a precision of $\pm 15\,\%$.
A small number of electrons are also produced in the Dalitz decay of 
$\pi^0$ and $\eta$ mesons. The Monte Carlo prediction was used for
this rate, and a $\pm 20\,\%$ uncertainty used to assess the
systematic error, based on measurements of inclusive $\pi^0$ and
$\eta$ production in hadronic events \cite{pi0prod}.
The total systematic error on \rb\ from all the
background sources was found to be $\pm \syselidbg$.

\subsection{Muon identification}

The muon identification efficiency was studied using various control
samples in data and Monte Carlo \cite{muonid}.
The muon matching requirements were studied using
muon pairs from two-photon production, for muons in the range 2 to
6\,GeV, and muon pairs from $\zb\rightarrow\mu^+\mu^-$
events for muons above 30\,GeV. The efficiency of the $\dedx$
requirement was studied using $\rm K^0_s\rightarrow\pi^+\pi^-$ decays.
The Monte Carlo was found to model the matching and $\dedx$ requirement
efficiencies to relative precisions of $2.1\,\%$ and $2.2\,\%$
respectively, to give a total error on the muon identification
efficiency of $3\,\%$. This  corresponds to an error of 
$\pm \sysmuidef$ on \rb. 

The various sources of background muons
were studied using control samples from $\rm K^0_s\rightarrow\pi^+\pi^-$ and
$\tau\rightarrow 3\pi$ decays, as in \cite{muonid}. The background in
the data was found to be a factor $1.13\pm 0.09$ higher than that in 
the Monte Carlo, and the fake probabilities per track were
corrected accordingly. The resulting uncertainty on \rb\ is $\pm \sysmuidbg$.

\subsection{Simulation input parameters}\label{ss:esimi}

The charm and light quark efficiencies are sensitive to the following
simulation input parameters:

\begin{description}

\item[Charm quark fragmentation:]
The charm tagging efficiency \ec\ increases with the scaled
energy $x_E$ of the weakly decaying charm hadron. The mean scaled
energy \meanxe\ of charm hadrons produced in \ccbar\ events at 
LEP has been measured by ALEPH, DELPHI and
OPAL \cite{bdfrag,dboth,dfrag}. These measurements have been averaged by 
the LEP electroweak working group to give a value of 
$\meanxe =0.484\pm 0.008$ for weakly decaying charm hadrons \cite{hfew}.
The effect of this uncertainty was assessed by reweighting 
events in the Monte Carlo changing \meanxe\ within this range. The
fragmentation functions of Peterson \cite{fpeter}, Collins and Spiller
\cite{fcolspil}, Kartvelishvili \cite{fkart} and the Lund group \cite{flund}
were used as models to determine these event
weights. The largest variation was found using the model of
Collins and Spiller, and the resulting variation in \rb\ of $\pm \syscqfrag$
was assigned as a systematic error. The charm fragmentation also
affects the tagging efficiency via the number of tracks produced in
the fragmentation process, but this effect was found to be much smaller than
the direct energy dependence and was neglected.

\item[Charm hadron production fractions:]
Because of the different charm hadron lifetimes and decay modes,
the vertex tagging efficiency in \ccbar\ events depends on the
mixture of weakly decaying charm hadrons.
The tagging efficiency
for $\rm D^+$ mesons is approximately three times that for $\rm D^0$,
whilst that for $\rm D_s^+$ mesons is approximately 15\%
higher than $\rm D^0$  and that for $\rm \Lambda^+_c$ is only
15\% of that for $\rm D^0$. The fractions of $\rm D^+$, $\rm D^0$,
$\rm D_s^+$ and $\rm\Lambda_c^+$ were varied according to the
production fractions measured at LEP \cite{dboth,dprod} as averaged by
the LEP electroweak working group \cite{hfew}.
The contribution from $\rm\Lambda_c^+$ was scaled by $1.15\pm 0.05$ to
account for other weakly decaying charm baryons. The errors were
combined taking their correlations into account to give an error on
\rb\ of $\pm \syschprov$. The dependence of the tagging efficiency on
the fraction of weakly decaying charm hadrons produced via the decay of excited
charm states ($\rm D^*$ and $\rm D^{**}$) was found to be negligible.

The systematic error on the lepton tagging efficiency in \ccbar\
events is derived from 
the inclusive charm semileptonic branching fraction, which is 
in turn derived from the
individual charm hadron semileptonic branching fractions (see below).
This introduces an additional dependence on the charm hadron
production fractions which is correlated with that from the vertex
tag, taking the total error on \rb\ from this source to $\pm \syschprod$.

\item[Charm hadron lifetimes:]
The lifetimes of the weakly decaying charm hadrons were varied
separately within
the errors quoted by the Particle Data Group \cite{pdg98}. Their
contributions to the error on \rb\ were added in quadrature to give an
error on \rb\ of $\pm \syschlife$.

\item[Charm hadron charged decay multiplicity:]
The  reconstructed secondary vertex track multiplicity $N_s$ 
is required to be at least three, and is also used as an input to
the vertex tag neural network. The tagging efficiency in \ccbar\
events therefore depends strongly on the charged track multiplicity of
charm hadron decays. The average charged track multiplicity of 
$\rm D^+$, $\rm D^0$ and  $\rm D_s^+$ decays (including the charged
decay products of any produced $\rm K^0_s$ mesons) has been measured by
MARK~III \cite{mark3}. The average multiplicity in Monte Carlo events was
varied within the range given by MARK III using several different
reweighting schemes, keeping the inclusive branching ratios to $\rm
K^0$ and $\Lambda$ constant for each charm hadron. For charm baryons,
for which no measurements are available, a variation of $\pm 0.5$ was
taken. The variations for each charm hadron species were combined in
quadrature to give an overall error on \rb\ of $\pm \syschmult$.

\item[Charm hadron neutral decay multiplicity:]
The charm tagging efficiency in the Monte Carlo is also observed to
depend on the number of $\pi^0$ mesons produced in the decay, even at fixed
charged decay multiplicity, as the number of $\pi^0$ mesons produced affects
the amount of energy and transverse momentum  available for the
charged decay products. The average $\pi^0$ multiplicity in $\rm D^+$,
$\rm D^0$ and $\rm D^+_s$ decays has been measured by MARK III
\cite{mark3}, but with large errors. However, for the $\rm D^+$ 
$(64\pm 5)\,\%$, and for the $\rm D^0$ $(76.2\pm 3.5)\,\%$, of the total decay
width is to known exclusive final states, most of which have well
measured branching fractions and low $\pi^0$ multiplicity \cite{pdg96}.

For the $\rm D^+$ and $\rm D^0$, the error is assessed by varying the
Monte Carlo $\pi^0$ multiplicity in the unmeasured decay
final states so as to reproduce the total variation in $\pi^0$ 
multiplicity allowed by the MARK III measurements, and adding a small 
contribution due to $\pi^0$ multiplicity variation allowed by the
measurement errors on the exclusive final state branching
fractions. Since most of the decay modes with high tagging
efficiency have well measured branching fractions, and the tagging
efficiency for the unmeasured decay modes is lower than average, this
procedure leads to total errors which are about half the
size of those obtained by
simply reweighting all decay final states to vary the $\pi^0$
multiplicity inclusively. The charged multiplicity and inclusive
branching ratios to $\rm K^0$ and $\Lambda$ were held constant.
The results of this procedure are
systematic errors on \rb\ of $\pm 0.00016$ for the $\rm D^+$ and 
$\pm 0.00021$ for the $\rm D^0$. For the $\rm D^+_s$, where only
$(18\pm 4)\,\%$ of the total decay width is measured, the inclusive
reweighting technique is used, leading to an error on \rb\
of $\pm 0.00016$. The error due to charm baryons is negligible, due to
their low production fraction and tagging efficiency. These errors are
combined in quadrature to give a total error on \rb\ of $\pm \syschneut$.

\item[Charm hadron to $\rm\bf K^0$ branching fraction:] The inclusive
  branching ratios $B(\rm D^+\rightarrow K^0,\overline{K^0}+X)$
$B(\rm D^0\rightarrow K^0,\overline{K^0}+X)$,
$B(\rm D^+_s\rightarrow K^0,\overline{K^0}+X)$ and
$B(\rm\Lambda^+_c\rightarrow \Lambda+X)$ were varied independently
  within the errors given by the Particle Data Group \cite{pdg98}, leaving
  the decay charged multiplicity distribution unaltered. The resulting
  variations in \rb\ were combined in quadrature to give a systematic
  error on \rb\ of $\pm \syschklam$.

\item[Charm semileptonic branching fraction:]
For semileptonic decays of charmed hadrons, an average branching
fraction of $(9.3\pm 0.5)\,\%$ was used. This value was obtained from
two sources: the direct charm semileptonic branching fraction of
$(9.5\pm 0.9)\,\%$ measured at $\sqrt{s}=10\rm\,GeV$ \cite{clargus},
and individual charm hadron branching fractions combined with the
production fractions discussed above. The charm hadron semileptonic 
branching fractions were obtained from the branching fraction 
$\rm D^0\rightarrow e^+ X$ of $(6.75\pm 0.29)\,\%$ \cite{pdg98},
together with the measured charm hadron lifetimes \cite{pdg98} and
assumptions of lepton universality and equal semileptonic widths of
all charm hadrons. This procedure gave  a charm semileptonic branching
fraction of $(9.1\pm 0.6)\,\%$, which was averaged with the direct 
measurement. The
direct measurement and $\rm D^0\rightarrow e^+$ branching ratio uncertainties
give a systematic error of $\pm \syscsemil$ on \rb\, and an additional
error of $\pm \syschprol$ results from the uncertainty in the 
charm hadron production fractions.

\item[Charm semileptonic decay modelling:]
The momentum spectra of the leptons in the rest frame of the decaying
charmed hadrons were modified according to the refined free-quark model
of Altarelli~\etal~\cite{accm}.
The two parameters of the model, $m_s$ and $p_F$, were chosen
to be $0.001\rm\,GeV$ and $0.467\rm\,GeV$, respectively, as given by a
fit to DELCO~\cite{delco} and MARK~III~\cite{mark3lept} data
performed by the LEP electroweak  working group.
Two sets of alternative values of the parameters,
$m_s=0.001\rm\,GeV$, $p_F=0.353\rm\,GeV$
and $m_s=0.153\rm\,GeV$, $p_F=0.467\rm\,GeV$,
corresponding to the variation allowed by the fit,
were used to estimate the systematic error on \rb\ of $\pm \syscsdmod$.

\item[Heavy quark production from gluon splitting:]
The production of heavy quark pairs via the processes 
$\rm g\rightarrow\ccbar$ and $\rm g\rightarrow\bbbar$ increases the
tagging efficiency in charm and light quark events. The rate of 
$\rm g\rightarrow\ccbar$ per multihadronic event has been measured by
OPAL to be $(2.38\pm 0.48)\times 10^{-2}$ \cite{opalgcc}, consistent
with perturbative QCD calculations \cite{theogcc}. The rate of 
$\rm g\rightarrow\bbbar$ has been measured by ALEPH and DELPHI
\cite{adgbb} and has been averaged to give a value of 
$(2.56\pm 0.67)\times 10^{-3}$ \cite{hfew}, also consistent with 
perturbative QCD calculations.
The Monte Carlo rates were adjusted to these central
values, and the uncertainties in the rates lead to errors on
\rb\ of $\pm \sysgsplcc$ from $\rm g\rightarrow\ccbar$ and 
$\pm \sysgsplbb$ from $\rm g\rightarrow\bbbar$.

\item[Inclusive $\rm\bf K^0$ and hyperon production:]
The total production rates of $\rm K^0$, $\Lambda$ and other weakly
decaying hyperons in the Monte Carlo were adjusted to agree with the
values measured by OPAL \cite{opalklh}. The rates were varied by
$\pm 3.4\%$, $\pm 6.5\%$ and $\pm 11.5\%$ respectively, corresponding
to the precision of the OPAL measurements combined with an additional 
uncertainty to take into account the extrapolation of the inclusive 
production rates to those for light quark events only.
The results of these
variations were combined in quadrature to give a systematic error on
\rb\ of $\pm \sysklhypr$. The dependence of the folded vertex 
tagging efficiencies on the number of tracks produced in light quark 
events was found to be negligible.

\end{description}
The finite number of charm and light quark Monte Carlo events contributes an
additional error of $\pm \sysmcstat$ on \rb.

\section{Systematic errors: efficiency correlation}\label{s:systc}

The tagging efficiency correlation \cb\ (as defined in 
section~\ref{s:anna}) is determined from a sample
of 7~million simulated \bbbar\ events, with small corrections for the
differing detector performance in each year. To evaluate the
systematic error on this quantity, 
three classes of effect that can give rise to an
efficiency correlation are considered: (1) kinematic
correlations due to final state gluon radiation, (2) geometrical
correlations due to detector non-uniformities, and (3) correlations
coming from the determination of the primary vertex position. From
equation~\ref{e:rbsens} it can be seen that the fractional error on
the correlation \cb\ contributes directly to the fractional error on
\rb, so an accurate evaluation of \cb\ is essential.

In general, correlations arise when the tagging efficiency \eb\
depends on a variable or variables $x$, and the values of $x$ are
correlated between the two hemispheres of the event. The resulting
tagging efficiency correlation \cb\ can be evaluated as:
\[
\cb = \frac{\mean{\eb(x)\eb(\overline{x})}}
{\mean{\eb(x)}\mean{\eb(\overline{x})}}
\]
where $x$ and $\overline{x}$ are the values of $x$ in the two
hemispheres of the event, and the average is taken over all \bbbar\
events in the sample. Such calculations, with the efficiencies
evaluated in small bins of $x$, are used frequently in the
correlation studies presented here.

The total correlation estimates and errors are
summarised in Table~\ref{t:cbsum}. The overall error on \rb\
resulting from uncertainties in the correlation values is 
$\pm \syscrltot$, corresponding to a relative uncertainty of $0.30\,\%$.
The systematic errors are evaluated by studying each
component of the correlation separately. It is then checked that the
total Monte Carlo correlation is reproduced by the sum of these
components. The dependence of the Monte Carlo correlation
on uncertainties in the detector simulation and physics modelling
is also taken into account. This method differs from that used in
\cite{pr188}, where only relatively small Monte Carlo samples were
available, and the overall correlation was determined by adding the
estimates of each separate correlation component.
Interdependence between the correlation components was treated 
in \cite{pr188} as an
additional source of systematic error, but this is now accounted for by
taking the correlation value directly from the Monte Carlo. 

\begin{table}
\centering 

\begin{tabular}{ll|rr}\hline \hline
Uncertainty on  $\cb-1$ (\%)& & Vertex & Combined \\ \hline
Kinematic & Same hemisphere events       & $0.08$ & $0.06$ \\
& Momentum correlation          & $0.33$ & $0.22$ \\ \hline
Geometrical & Systematic error       & $0.03$ & $0.02$ \\
            & Data statistics 1992 & $0.14$ & $0.12$ \\
            & Data statistics 1993 & $0.17$  & $0.11$ \\
            & Data statistics 1994 & $0.09$ & $0.06$ \\
            & Data statistics 1995 & $0.13$ & $0.09$ \\ \hline
Primary vertex  &   &  $0.02$ & $0.02$ \\ \hline
Simulation & Detector resolution    & $0.01$ & $0.02$ \\
& Beam spot size         & $0.04$ & $0.06$ \\
& b quark fragmentation  & $0.09$ & $0.08$ \\
& b hadron lifetime      & $0.04$ & $0.03$ \\
& b decay multiplicity   & $0.01$ & $0.01$ \\
\hline
Total systematic error & & $0.36$ & $0.25$ \\
\hline
Monte Carlo statistics & & $0.20$ & $0.17$ \\
\hline \hline
Total Correlation $\cb-1$ (\%)
& 1992 & $0.84\pm 0.42$ & $1.01\pm 0.33$ \\
& 1993 & $1.10\pm 0.44$ & $1.13\pm 0.33$ \\
& 1994 & $0.77\pm 0.41$ & $0.92\pm 0.31$ \\
& 1995 & $0.99\pm 0.42$ & $0.95\pm 0.32$ \\
\hline
\end{tabular}
\caption{\label{t:cbsum} Errors on the hemisphere
  efficiency correlation $\cb-1$ for the vertex and combined tags, 
  together with the correlation values
  and associated total errors used for each year of the data. The
  latter are obtained using the geometrical correlation values given
  in Table~\ref{t:geoc}.}
\end{table}

The evaluation of the kinematic correlation is described in detail
in Sections~\ref{ss:sameh} and~\ref{ss:cbmom}, and various comparisons
between data and Monte Carlo are discussed in Sections~\ref{ss:cbdat}
and~\ref{ss:cbep}. The
geometrical correlation, and the associated corrections to the Monte
Carlo correlation value, are described in Section~\ref{ss:geoc}.
Primary vertex effects are discussed in Section~\ref{ss:cpvtx} and
the detector simulation and physics modelling systematic errors in
Section~\ref{ss:csim}. The completeness test, that the total Monte
Carlo correlation is consistent with the sum of the various
components, is described in Section~\ref{ss:compl}.
Throughout this section, correlation values and
uncertainties are given both for the vertex tag alone, and for the
combined vertex and lepton tags.

\subsection{Kinematic Correlation---same-hemisphere events}\label{ss:sameh}

If an energetic gluon is radiated in a \bbbar\ event, it may cause
both b hadrons to recoil into the same thrust hemisphere. 
The tagging efficiency
for such hemispheres is lower than that for the average b hemisphere,
since the b hadrons have much lower momentum, and 
the double tagging efficiency for these events is much lower 
than for normal \bbbar\ events since one hemisphere contains no b
hadrons. These events therefore introduce a small
hemisphere efficiency correlation, which can be
calculated in the Monte Carlo from the number of such events and their
hemisphere and double tagging efficiencies.
The Monte Carlo predicts that 1.21\,\% of \bbbar\ events
passing the event selection have both b hadrons in the same
hemisphere. However, such events constitute only 0.59\,\% of the tagged
hemispheres and 0.003\,\% of the double tagged events for the vertex
tag, and only 0.62\,\% and 0.014\,\% for the combined tag. They
contribute $(0.02\pm 0.02)\,\%$ to the tagging efficiency correlation for the
vertex tag and $(-0.03\pm 0.02)\,\%$ for the combined tag, where the
errors are due to Monte Carlo statistics.

The number of same hemisphere events was compared in data and Monte
Carlo by looking for events with two vertex tags in the same
hemisphere, each passing a very loose neural network output cut of $B>0.5$.
In the Monte Carlo, $36\,\%$ of these double tagged hemispheres come
from genuine
same-hemisphere events, $40\,\%$ from normal \bbbar\ events with a
mis-reconstructed vertex, and $23\,\%$ from charm and light quark
events. The different contributions were statistically separated using the
three-dimensional angle \phiv\ between the momentum vectors of
the two vertices; the same-hemisphere \bbbar\ events have a relatively
broad \phiv\ distribution and the remaining contributions 
are more concentrated at low values of \phiv. The 
distribution of \phiv\ for the double tagged hemispheres in data and
Monte Carlo is shown in Figure~\ref{f:csameh}.
The \phiv\ distribution in the data was fitted to the sum of the Monte
Carlo distributions, allowing the rate  of the Monte Carlo 
same hemisphere \bbbar\ events to vary. The rate of genuine same hemisphere
events was found to be consistent between data and Monte Carlo with
a statistical uncertainty of $\pm 40\,\%$. The systematic error due to
uncertainties in the momentum distribution of the b hadrons was
evaluated by comparing the predictions of JETSET 7.4 and 
HERWIG~5.9 \cite{herwig}. This gives  a 7\,\% relative uncertainty on the
vertex tagging efficiency for same hemisphere events, and a 5\,\%
uncertainty for the combined tag. The resulting total 
uncertainty on the size of the same hemisphere event
correlation is $\pm 0.08\,\%$ for the vertex tag and $\pm 0.06\,\%$
for the combined tag. The tagging efficiencies are also sensitive to
the modelling of b hadron decay in the Monte Carlo, but this is
addressed in Section~\ref{ss:csim}.

\epostfig{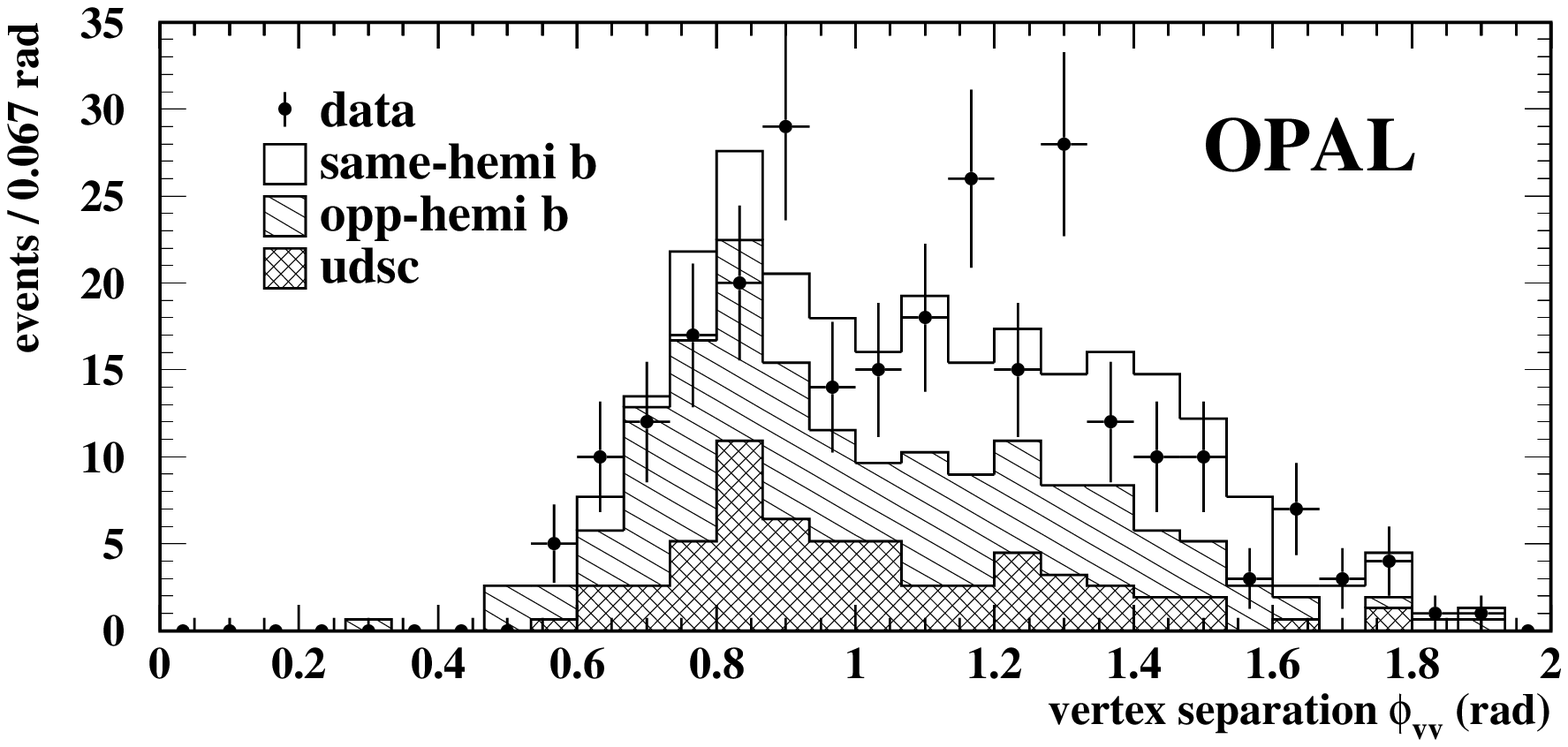}{f:csameh}{Distribution of the separation
angle \phiv\ between two vertices loosely tagged in the same hemisphere
for genuine same-hemisphere \bbbar\ events, normal opposite 
hemisphere \bbbar\ events,
charm and light quark events and data.}

\subsection{Kinematic Correlation---momentum and fragmentation}\label{ss:cbmom}

After removing the same-hemisphere events in the Monte Carlo, 
each hemisphere of a
\bbbar\ event contains one b hadron. Kinematic variables, such as 
the momenta of the two b hadrons \pb\ and \pbbar,
are correlated between the two hemispheres due to final state gluon radiation.
Since the b-tagging efficiency depends on these kinematic variables, 
such correlations can produce a tagging efficiency correlation.

The b-tagging efficiency depends strongly on the b hadron
momentum in the same hemisphere, since high-momentum 
b hadrons are likely to travel further before decaying,
and decay producing higher momentum tracks which are measured with better
resolution. Both of these effects produce a 
 more easily resolvable secondary vertex. The efficiency is shown as 
a function of $\xb=\pb/E_{\rm beam}$ for
the combined vertex and lepton tag in Figure~\ref{f:corlint}(a).
However, the b-tagging efficiency also depends on properties of 
the fragmentation tracks in the same hemisphere, for example their
number, momentum and angular distribution,
which influence the reconstruction of the primary and secondary vertices.
The dependence of the b-tagging efficiency on the number of
fragmentation tracks \nfrag\ is shown in Figure~\ref{f:corlint}(b). 
Although \nfrag\ is negatively correlated with \pb\, there is a 
dependence even at constant \pb. As the number of 
fragmentation tracks in a hemisphere increases,
the vertex tagging algorithm becomes increasingly  likely to reconstruct 
the primary rather than the secondary vertex, and is therefore less likely 
to give a b-tag.

\epostfig{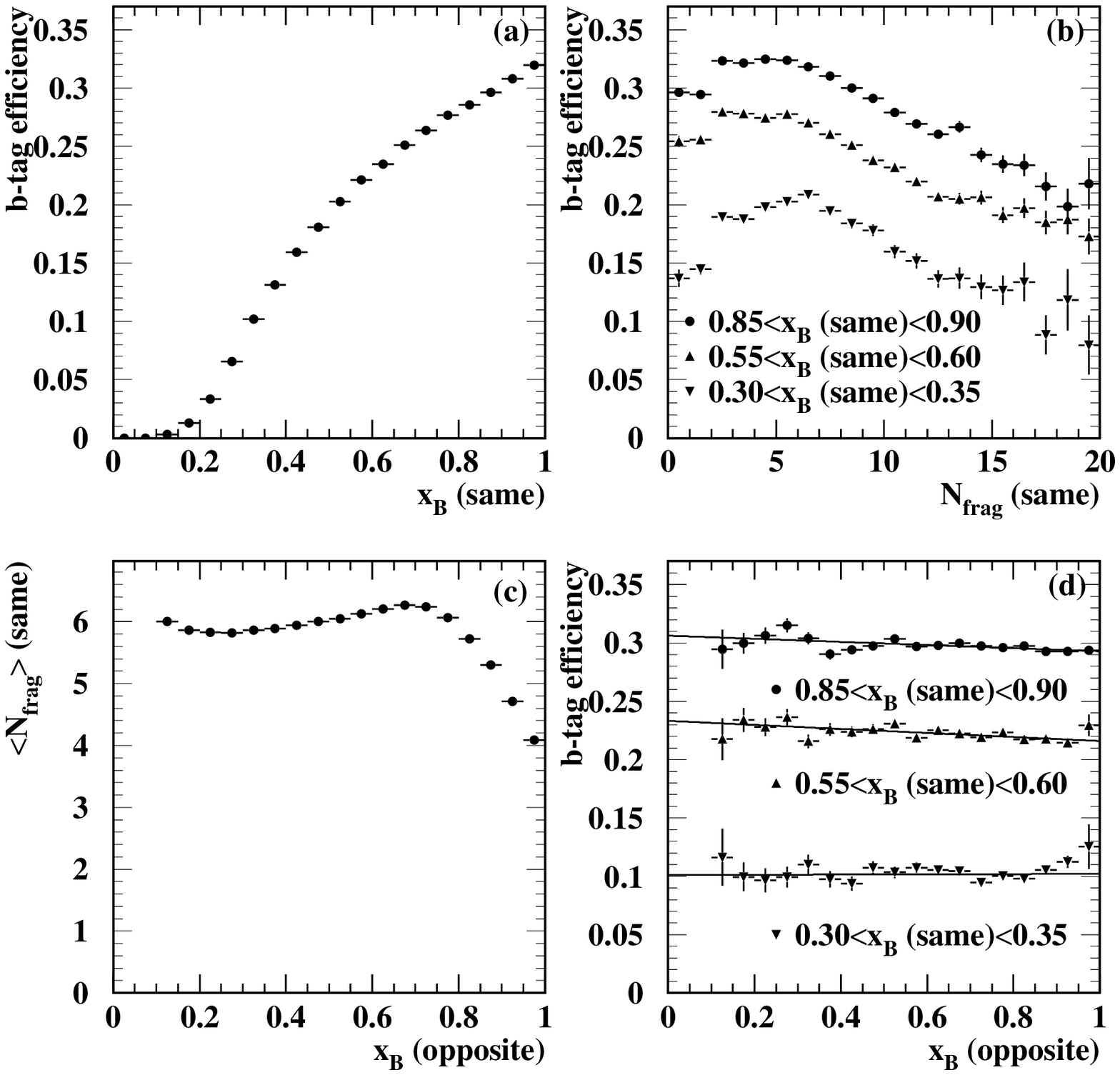}{f:corlint}{Origins of kinematic correlations
  in Monte Carlo \bbbar\ events: (a) Dependence of b-tagging
  efficiency \eb\ on the scaled b hadron momentum 
  $\xb=\pb/E_{\rm beam}$ in the same
  hemisphere; (b) dependence of \eb\ on the number of
  fragmentation tracks \nfrag\ in the same hemisphere, at various
  values of \xb\ in the same hemisphere; (c) variation in the
  mean number of fragmentation tracks $\mean{\nfrag}$ with \xb\
  in the opposite hemisphere; (d) dependence of \eb\ on \xb\ in
  the opposite hemisphere, at various values of \xb\ in the
  same hemisphere.}

Final state gluon radiation into one hemisphere decreases the momentum
of the b hadron in that hemisphere, and increases the number of
fragmentation tracks. It will also increase the hemisphere mass 
$\mh=\sqrt{\eh^2-\ph^2}$ where \eh\ and \ph\ are the total energy and
momentum in the hemisphere. Since the total momenta of the two
hemispheres must balance, this will also reduce the total momentum,
and hence the momentum available to the b hadron, in the opposite 
hemisphere of the event. Therefore the momentum \pb\ of the b
hadron in one hemisphere is correlated both with the momentum \pbbar\
of the b hadron and the number of fragmentation tracks \nfragbar\ in
the opposite hemisphere. This can be seen in
Figure~\ref{f:corlint}(c), where the Monte Carlo dependence of \nfrag\
in one hemisphere on \pb\ in the other hemisphere is shown.
At moderate values 
of \pbbar, $\mean{\nfrag}$ increases with increasing \pbbar\, 
and it then decreases again at very high values of \pbbar.

As a result of these correlations, the b-tagging efficiency is found
to depend not only on the same hemisphere b hadron momentum \pb, but also
weakly on the opposite hemisphere b hadron momentum \pbbar, as shown
in Figure~\ref{f:corlint}(d). 
This effect was also seen with the much simpler vertex tagging 
algorithm used in \cite{pr188}.
The resulting kinematic correlation can be calculated by 
parameterising the b-tagging efficiency as a function of both \pb\ and 
\pbbar, and is given by:
\begin{equation}\label{e:cbpp}
\cbpp = \frac{\mean{\eb(\pb,\pbbar)\eb(\pbbar,\pb)}}
{\mean{\eb(\pb,\pbbar)}\mean{\eb(\pbbar,\pb)}} ,
\end{equation}
where $\eb(p_1,p_2)$ gives the b-tagging efficiency for a hemisphere
containing a b hadron of momentum $p_1$, the other hemisphere of the
event containing a b hadron of momentum $p_2$, and the averages are taken 
over all \bbbar\ events in the sample. The Monte Carlo
predicts that the kinematic efficiency correlation is 
$\cbpp-1=(0.04\pm 0.05)\,\%$ for the vertex tag and 
$\cbpp-1=(0.06\pm 0.03)\,\%$ for the combined tag.

The size of the underlying momentum correlation between \pb\ and 
\pbbar\ is given by:
\[
\cpb = \frac{\mean{\pb\ \pbbar}}{\mean{\pb}\mean{\pbbar}} ,
\]
and the size of the resulting efficiency correlation, if the b-tagging
efficiency depended only on the same hemisphere b hadron momentum,
would be:
\begin{equation}\label{e:cbp}
\cbp = \frac{\mean{\eb(\pb)\ \eb(\pbbar)}}
{\mean{\eb(\pb)}\mean{\eb(\pbbar)}} ,
\end{equation}
The Monte Carlo predicts a b hadron momentum correlation of 
$\cpb-1=(0.80\pm 0.01)\,\%$, in agreement with perturbative QCD
calculations \cite{nasoncb}. This would result in a tagging efficiency
correlation of $\cbp-1=(0.87\pm 0.03)\,\%$ for the vertex tag and 
$\cbp-1=(0.55\pm 0.03)\,\%$ for the combined vertex and lepton
tag. The additional dependence of the tagging efficiency on \pbbar\
via the fragmentation tracks reduces this correlation, 
to give an overall effect which
is approximately zero. However,  both the size of the momentum 
correlation \cpb\ and the separate correlation between \nfrag\ and \pbbar\ are 
tested when calculating the systematic error. This is discussed in 
Section~\ref{ss:cbdat} below.

The momentum correlation has been treated above in terms of \pb\
and \nfrag, quantities on which the tagging efficiency 
depends strongly, but whose hemisphere correlations are weak and
difficult to measure. A  different and complementary approach is provided by
considering the total energy \eh\ and momentum \ph\ in each hemisphere, which
are completely correlated by overall energy and momentum conservation
in the event, and are affected by final state gluon
radiation. These variables should give a good description of the
kinematic correlation resulting purely from energy and momentum 
conservation. The variables \eh\
and \ph\ are relatively easy to measure in the data, and the two
hemispheres are related by
$\eh+\ehbar=2\,E_{\rm beam}$ and $\ph=\phbar$.
However, the dependence of the tagging efficiency on these
quantities is weaker than that on \pb\ and \nfrag.

In this approach, the kinematic correlation resulting from energy and
momentum conservation may be calculated by parameterising the tagging
efficiencies in terms of \eh\ and \ph :
\begin{equation}\label{e:cbep}
\cbep = \frac{\mean{\eb(\eh,\ph)\eb(\ehbar,\phbar)}}
{\mean{\eb(\eh,\ph)}\mean{\eb(\ehbar,\phbar)}} ,
\end{equation}
The Monte Carlo
predicts that $\cbep-1=(-0.27\pm 0.04)\,\%$ for the vertex tag and 
$(-0.14\pm 0.04)\,\%$ for the combined tag. If the efficiency is
parameterised in terms of the hemisphere momentum \ph\ alone, the 
correlation is $(1.34\pm 0.03)\,\%$ for the vertex tag and 
$(0.85\pm 0.03)\,\%$ for the combined tag, somewhat larger than the
values obtained for \cbp\ using the b hadron momentum \pb\ alone.

The kinematic correlation
$\cbep-1$ is significantly more negative than $\cbpp-1$.
To study whether significant additional
correlation, other than that caused by energy and momentum
conservation, is present in the Monte Carlo, the correlation \cpb\ between
\pb\ and \pbbar\ was evaluated in bins of \eh\ and \ph. For the
majority of the events, no significant additional correlation was
observed, but for events with low values of \ph, {\em i.e.} those with
broad high mass jets in both hemispheres, a significant positive
correlation between \pb\ and \pbbar\ was observed. This gives an extra
contribution of about $+0.15\,\%$ to the tagging efficiency
correlation, bringing the overall kinematic correlation closer to that
calculated from \pb\ and \pbbar\ above.

Both of the approaches described above were used to study correlations
in the data. The correlations between \pb\ and \pbbar, and between
\nfrag\ and \pbbar, were checked using estimators of these quantities 
in a sample of loose double tagged events (see Section~\ref{ss:cbdat}). 
The distributions of
\ph\ and \eh, and the dependence of the tagging rates on them were studied.
Monte Carlo events were
reweighted to make the various distributions the same as those in the
data, and the resulting changes in the correlation evaluated
(see Section~\ref{ss:cbep}). Since
all of the tests are somewhat indirect, and all the possible sources
of kinematic correlations have not been completely understood, the
largest systematic error resulting from any of the tests is used as
the overall systematic error on the kinematic correlation. As can be
seen from Table~\ref{t:cbsum}, the resulting systematic error is
similar in size to the uncertainty in the correlation from Monte Carlo
statistics.

\subsection{Kinematic Correlation tests---b momentum and fragmentation}
\label{ss:cbdat}

In order to study correlations in the data, a loose double tag was used 
to select a sample of almost pure \bbbar\ events. Events were selected
if each hemisphere contained a secondary vertex passing the neural network 
pre-selection described in Section~\ref{ss:svtx}, 
{\em i.e.\/} at least three tracks with a decay length 
significance of at least $|L/\sigma|>3$, or a lepton passing the
selection described in Section~\ref{s:lepton}. This selection has an
efficiency for \bbbar\ events of about $31\,\%$ and a b purity of about 95\,\%.

Reconstructed secondary vertices were then used to
form estimates \pv\ of the corresponding b hadron momentum. The
component of the b hadron's momentum associated with charged particles
was estimated by summing the momenta of all charged particles in the
jet, each weighted by $X$---the weight for the track to have come from
the b hadron decay described in Section~\ref{ss:vmass}. The neutral
momentum component was estimated using the energy of all the
electromagnetic calorimeter clusters in the jet, each weighted
as a function of the angle between the cluster and the jet
axis. If any charged tracks were associated to the cluster, their
energies  were first subtracted from the cluster energy, and the
cluster was only used if the remaining energy was greater than zero.
If the hemisphere was tagged only by a lepton, the lepton momentum was used
as the \pv\ estimator. The lepton momentum was scaled to give the same
mean value of \pv\ as that from reconstructed vertices.
The correlation of $\xv=\pv/E_{\rm beam}$ with 
$\xb=\pb/E_{\rm beam}$ is shown in Figure~\ref{f:corlest}(a),
and the distribution of \xv\ in data and Monte Carlo loose
double-tagged events is shown in Figure~\ref{f:corlest}(b).
The correlation coefficient
$\rho=(\mean{\pb\pv}-\mean{\pb}\mean{\pv})/(\sigma_{\pb}\sigma_{\pv})$
of the two quantities is 0.34, where $\sigma_{\pb}$ and $\sigma_{\pv}$
are the standard deviations of $\pb$ and $\pv$ respectively.

\epostfig{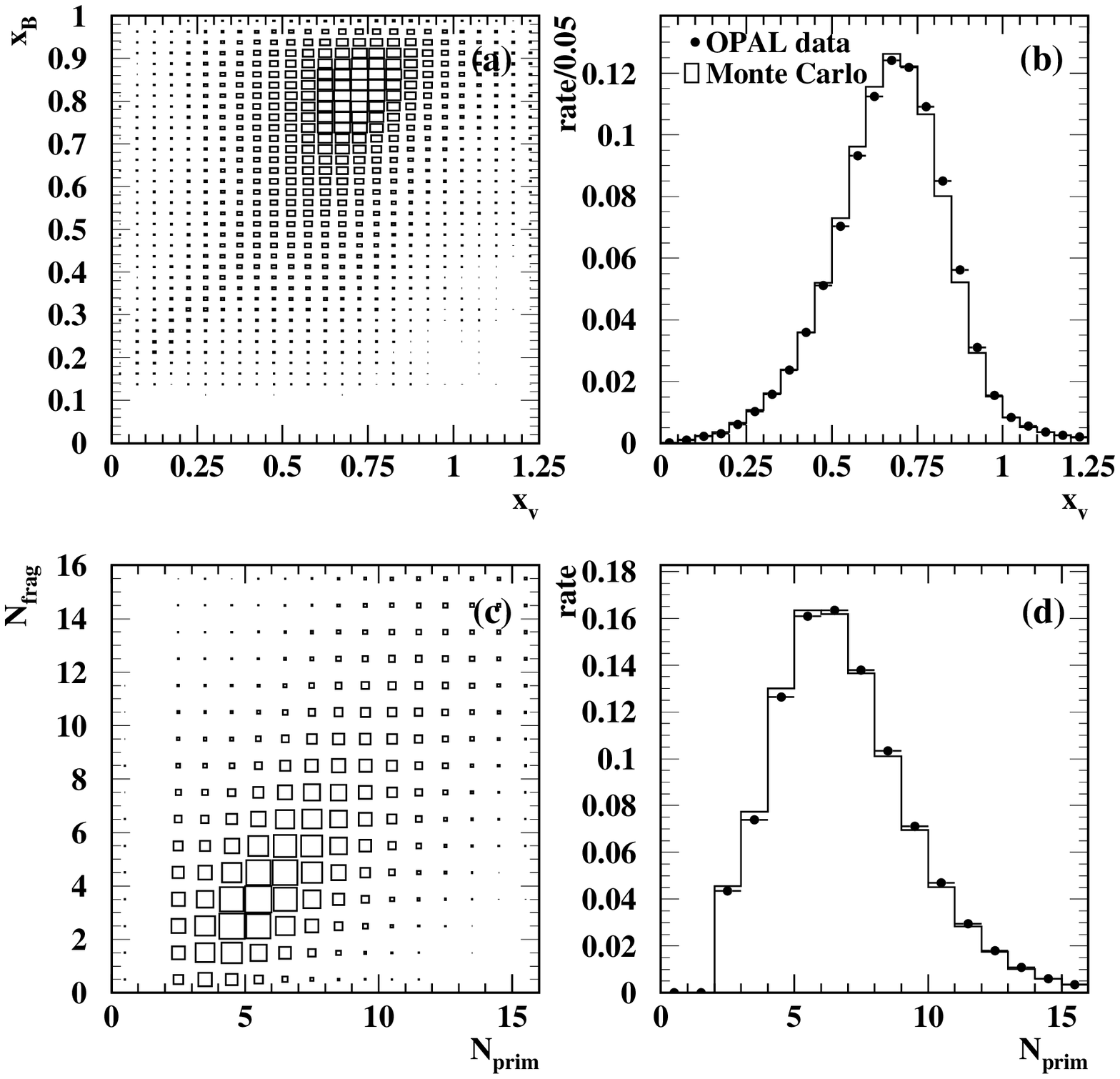}{f:corlest}{Estimators of of the scaled b hadron
  momentum \xb\ and number of fragmentation tracks \nfrag:
  (a) correlation of estimated \xb\ (\xv) with
true \xb\ (b) distributions of \xv\ in
data (points) and Monte Carlo (histogram), 
(c) correlation of estimated \nfrag\ (\nprim) and true \nfrag,
(d) distributions of \nprim\ in data and Monte Carlo.}

The number of tracks assigned to the hemisphere primary vertex \nprim\
was used as an estimator of the number of fragmentation tracks \nfrag\
in the hemisphere. The correlation of \nprim\ with \nfrag\ is shown in
Figure~\ref{f:corlest}(c), and the distribution of \nprim\ in data and
Monte Carlo in Figure~\ref{f:corlest}(d). The correlation
coefficient of \nprim\ and \nfrag\ is 0.68.

The correlation of the estimated b hadron momentum in one hemisphere,
\pv, with the estimated b hadron momentum in the other hemisphere,
\pvbar\ was measured to be $\cpv-1=(0.670\pm 0.023)\,\%$ in the loose
double tagged data sample, and $\cpv-1=(0.664\pm 0.015)\,\%$ in the
corresponding Monte Carlo sample. The effects of including the
same-hemisphere events and the non-\bbbar\ background were checked in
the Monte Carlo and found to be negligible. The Monte Carlo is
therefore seen to reproduce the strength of the correlation between
\pv\ and \pvbar\ seen in the data. Since \pv\ is only an estimate of
\pb\, the Monte Carlo was reweighted to change the true correlation
between \pb\ and \pbbar\, and the resulting changes in \cpv\ and the
efficiency correlation \cbp\ studied.  The statistical error on the
difference between \cpv\ in data and Monte Carlo corresponds to an
uncertainty of $\pm 0.27\,\%$ in \cbp\ for
the vertex tag and $\pm 0.19\,\%$ for the combined tag.

The size of the correlation between \nfrag\ and the opposite
hemisphere b hadron momentum was also studied using the estimators
described above. The average values of \nprim\ as a function of the
estimated opposite hemisphere b momentum are shown in
Figure~\ref{f:corldmc}(a) for the Monte Carlo and
Figure~\ref{f:corldmc}(b) for the data loose double tagged samples. A
complicated correlation is seen in both data and Monte Carlo,
 with the value of $\mean{\nprim}$
increasing in the region $0.2<x_{\rm v}<0.5$ and then decreasing again in
the region $0.5<\xv<0.8$. Qualitatively similar behaviour is also seen 
in the variation of $\mean{\nfrag}$ with \xb\ shown in 
Figure~\ref{f:corlint}(c).

\epostfig{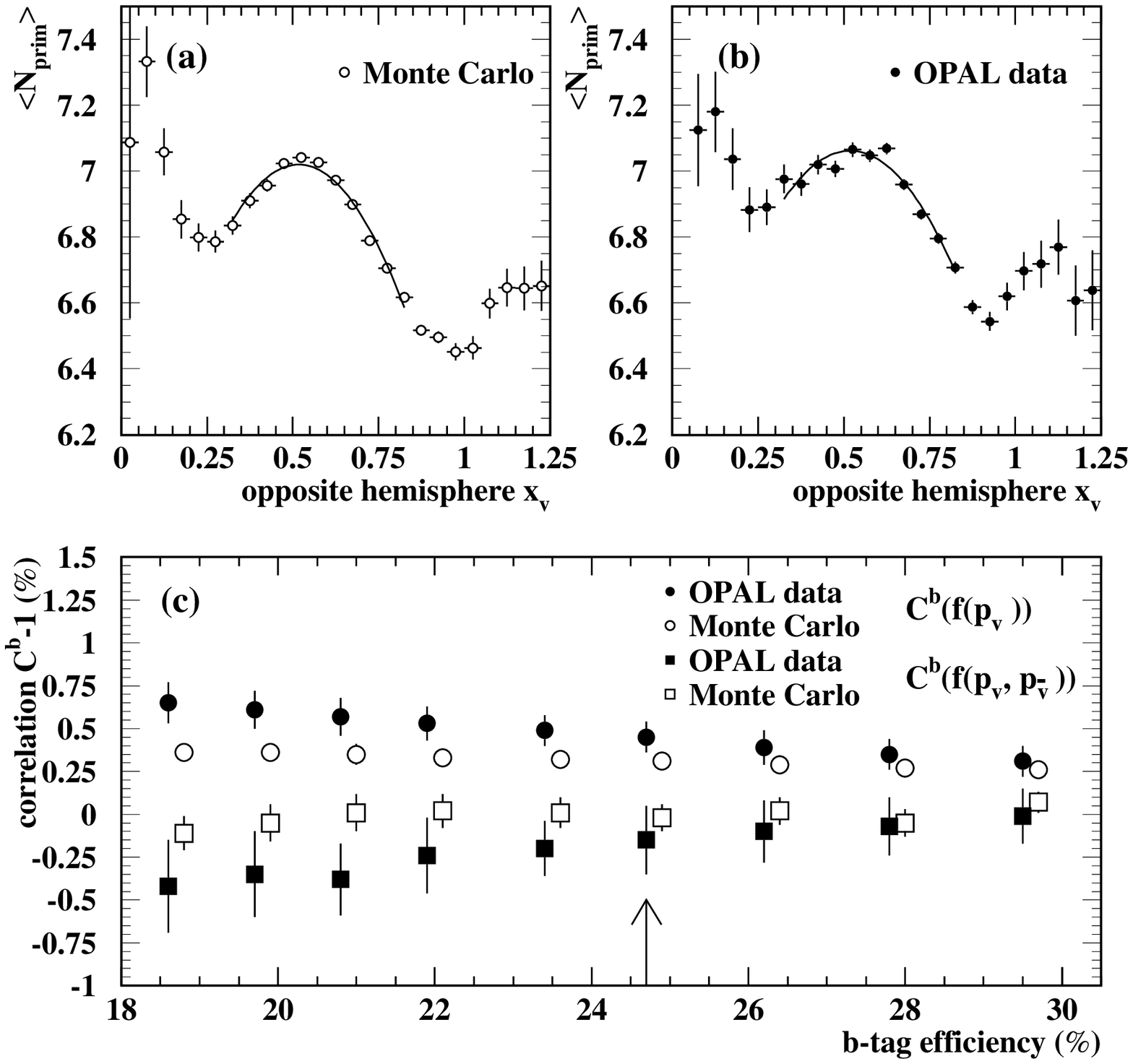}{f:corldmc}{Estimated $\mean{\nfrag}$ as a
  function of the estimated opposite hemisphere scaled momentum
  \xb, in loose double tagged Monte Carlo (a) and data (b)
  samples. The fits used to estimate the size of the correlation are
  shown. (c) Hemisphere tagging efficiency
  correlations $\cb-1$ calculated using the estimated b hadron
  momentum \pv, as a function of \pv\ only in the tagging hemisphere
  (circles), and in both the tagging and opposite hemispheres
  (squares). The data is shown by the filled symbols and the Monte
  Carlo by the open symbols. The errors are highly correlated between
  the points of different b-tagging efficiency, and the 
  analysis cut is shown by the arrow. The individual points are
  slightly displaced along the $x$-axis for clarity.}

Two methods were used to compare the strength of the $\nfrag,\pbbar$
correlations seen in data and Monte Carlo.
In the first method, the distributions were fitted to a
parabola in the region $0.3<\xvbar<0.8$ (where the bulk of the events
lie), and the curvatures compared.
The ratio between data and Monte Carlo curvatures was found to be
$0.85\pm 0.10$. In the second method, the root mean square deviation
of $\mean{\nprim}(\xvbar)$ from $\mean{\mean{\nprim}}$ was calculated,
weighted according to the distribution of \xvbar. Here, 
$\mean{\nprim}(\xvbar)$ is the average value of \nprim\ at a
particular value of \xvbar, and $\mean{\mean{\nprim}}$ is the average
of \nprim\ calculated over all events.
This method also takes into account the
events outside the region $0.3<\xvbar<0.8$, and gives a ratio between
data and Monte Carlo of 0.91. The Monte Carlo was then reweighted to
change the correlation between \nfrag\ and \pbbar, and the variation
of the difference $\cbpp-\cbp$ studied. The full variation between the
unweighted Monte Carlo sample and the sample reweighted to reduce the 
correlation between \nprim\ and \pvbar\ by 25\,\% was used to estimate
the systematic error on $\cbpp-\cbp$, giving uncertainties of 
$\pm 0.20\,\%$ and $\pm 0.12\,\%$ for the vertex and combined tags 
respectively.

The uncertainties on the separate components of the kinematic
correlation $\cbp$ and $\cbpp-\cbp$ were then added in quadrature to
give total uncertainties on the kinematic correlation of $0.33\,\%$ and
$0.22\,\%$ for the vertex and combined tags respectively. These
uncertainties are the largest from any of the correlation tests, and
set the size of the overall systematic error. The
uncertainty for the vertex tag alone is somewhat larger for two
reasons: the lepton tag has a weaker dependence on the
b hadron momentum and no dependence on the number of fragmentation
tracks; and the comparisons between data and Monte Carlo are dominated
by the statistical precision of the tests, which are improved by
adding the extra lepton tagged hemispheres.

As an alternative approach, the distributions of \pv\ and \nprim\ were
studied in data and Monte Carlo, and the influence of discrepancies in
these distributions on the kinematic correlation investigated. In this
approach, the decomposition of the kinematic correlation into
the $\cbp$ and $\cbpp-\cbp$ components was not used.
Event weights were calculated for the Monte Carlo to make the
distributions of \pv\ and \nprim\ match those in the data, and the resulting 
kinematic correlation calculated and compared with the value derived
using:
\[
\cbvv = \frac{\mean{\eb(\pv,\pvbar)\eb(\pvbar,\pv)}}
{\mean{\eb(\pv,\pvbar)}\mean{\eb(\pvbar,\pv)}}
\]
{\em i.e.\/} the same as equation~\ref{e:cbpp}, but with \pb\ and
\pbbar\ replaced by the estimators \pv\ and \pvbar\, and the averages
taken over the sample of loose double tagged events. Event weights
and efficiencies were calculated as a function of \pv\ and \pvbar,
\pv\ and \nprim, and \nprim\ and \nprimbar, and the largest changes
seen in \cbvv\ were $0.11\,\%$  and $0.07\,\%$ for the vertex and
combined tags respectively, well within the systematic errors assigned
for the kinematic correlation.

Finally, the hemisphere tagging efficiency 
relative to the loose double tag  pre-selection was calculated 
directly in the data and Monte Carlo, from the fraction of events
tagged as a function of \pv , or of \pv\ and \pvbar. The two
dimensional distribution of \pv\ and \pvbar\ can then be used to
derive an effective tagging efficiency correlation. The results of
this procedure are shown in Figure~\ref{f:corldmc}(c). The circles show
the correlation calculated using the momentum dependence of the
tagging efficiency in the tag hemisphere only:
\[
\cb (f(\pv))  = \frac{\mean{f(\pv) f(\pvbar)}}
{\mean{f(\pv)} \mean{f(\pvbar)}} ,
\]
where $f(\pv)$ is the fraction of hemispheres of estimated b
hadron momentum \pv\ that are tagged. The squares show the correlation
calculated using the momentum dependence of the tagging efficiency in
both hemispheres:
\[
\cb (f(\pv,\pvbar)) = \frac{\mean{f(\pv,\pvbar) f(\pvbar,\pv)}}
{\mean{f(\pv,\pvbar)} \mean{f(\pvbar,\pv)}} .
\]
In both cases, the correlations calculated from data and Monte Carlo 
are similar, and the increasing 
size of the correlation as the b-tagging efficiency is reduced is well
modelled. The correlation using the tagging hemisphere momentum
dependence alone is similar to that obtained in the Monte Carlo using the
true b hadron momentum. The correlation due to the momentum dependence
in both hemispheres is much closer to zero, again as in the Monte
Carlo using the true b hadron momentum, though the
statistical precision of this test is rather low.

\subsection{Kinematic Correlation tests---hemisphere energy and momentum}
\label{ss:cbep}

To study the correlations expressed in terms of hemisphere energy and
momentum, estimates
were formed of the total energy and momentum in each hemisphere using both
tracking and calorimeter information, and applying an algorithm to
correct for double counting of charged particles \cite{gcecode}.
These reconstructed energy and momentum distributions have widths of
about 7\,GeV and are dominated by the experimental resolution. Since
the hemisphere correlations of these quantities are known,
the resolution of the estimators can be improved using information
from both hemispheres and the beam energy constraint. This is done 
by calculating the two hemisphere masses 
\mh\ and \mhbar, using the beam energy to scale the total observed
energy:
\[
\mh = \frac{2E_{\rm beam}}{\eh +\ehbar}\sqrt{\eh^2-\ph^2}
\]
and then recalculating the hemisphere energies and momenta,
using the two hemisphere masses, the known beam energy and two-body
decay kinematics. The
correlation of the reconstructed hemisphere mass with the true mass in
Monte Carlo \bbbar\ events is 
shown in Figure~\ref{f:cep1}(a), and the reconstructed mass
distributions for the loose
double tagged data and Monte Carlo samples in Figure~\ref{f:cep1}(b). 

The resulting 
distributions of hemisphere energies and momenta are shown in
Figure~\ref{f:cep1}(c) and (d). The resolutions of the estimators of
hemisphere energy and momentum are 0.82\,GeV and 0.68\,GeV
respectively, where the true quantities include unmeasured particles
such as neutrinos. Some discrepancies are seen between the data and
Monte Carlo distributions, which are addressed in the systematic error
evaluation below.

\epostfig{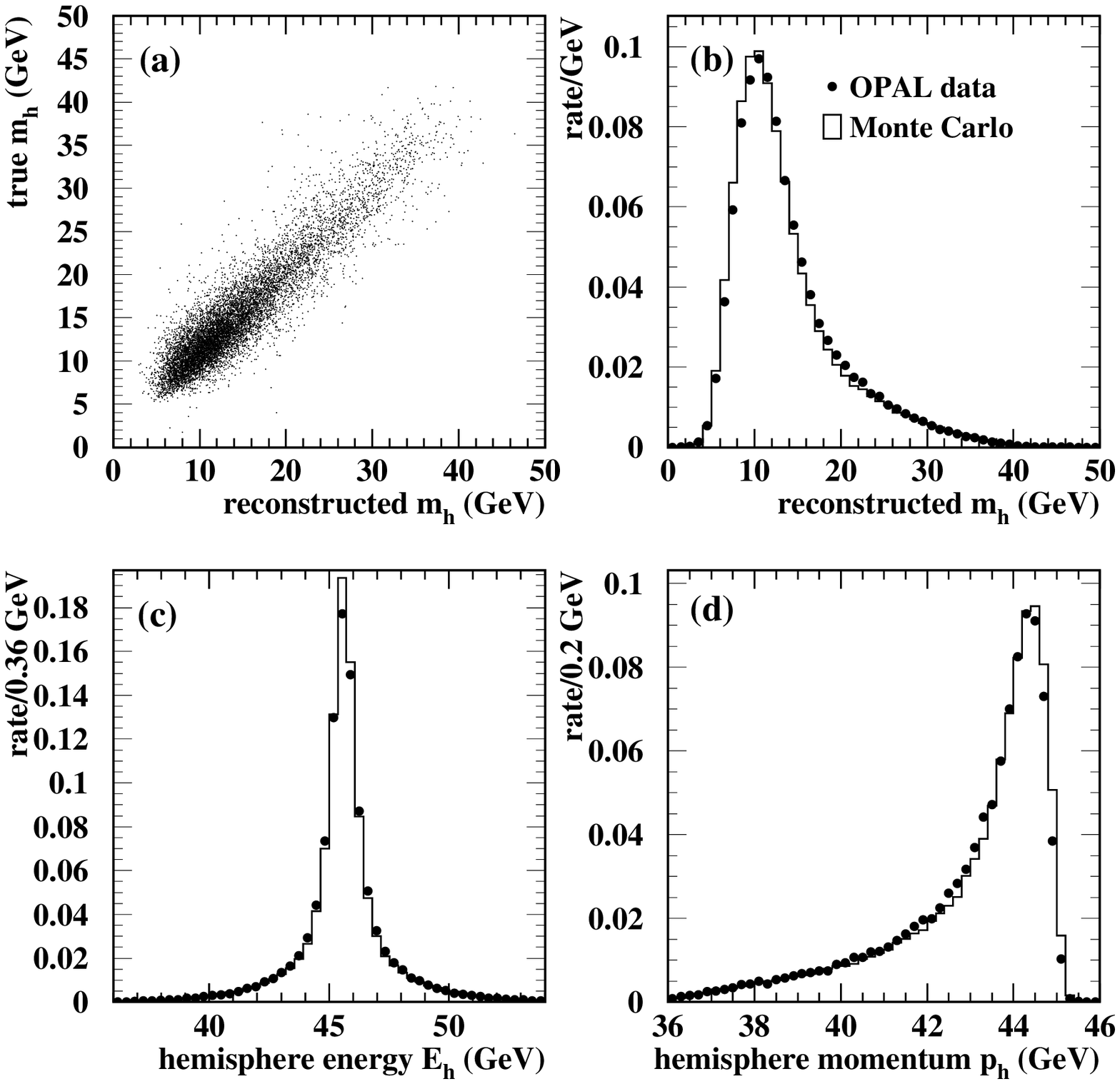}{f:cep1}{(a) Correlation of reconstructed and true
hemisphere masses \mh; (b--d) Distributions of reconstructed hemisphere 
mass, energy and momentum in loose double tagged events in data
(points) and Monte Carlo (histogram).}

The systematic error on \cbep\ is evaluated in two steps. First, the
two-dimensional distribution of hemisphere energies \eh\ and momenta \ph\ is
reweighted in the loose double tagged Monte Carlo to match that 
of the data. This gives a
shift in \cbep\ of $(+0.08\pm 0.01)\,\%$ for the vertex tag and 
$(+0.04\pm 0.01)\,\%$ for the combined tag, where the errors are due
to data and Monte Carlo statistics.
Secondly, the modelling of the tag efficiency as a
function of \eh\ and \ph\ is checked, by studying the fraction of
tagged hemispheres as a function of \eh\ and \ph\ in data and Monte
Carlo, as shown in Figure~\ref{f:cep2}. Reweighting the Monte Carlo
to make the two dimensional  distribution of the tag fraction as
a function of \eh\ and \ph\ agree results in a change in \cbep\ of
$(+0.03\pm 0.15)\,\%$ for the vertex tag and $(+0.03\pm 0.13)\,\%$ 
for the combined tag, where the errors are statistical. These changes
are all small, and well within the overall systematic errors assigned
for the kinematic correlation.

\epostfig{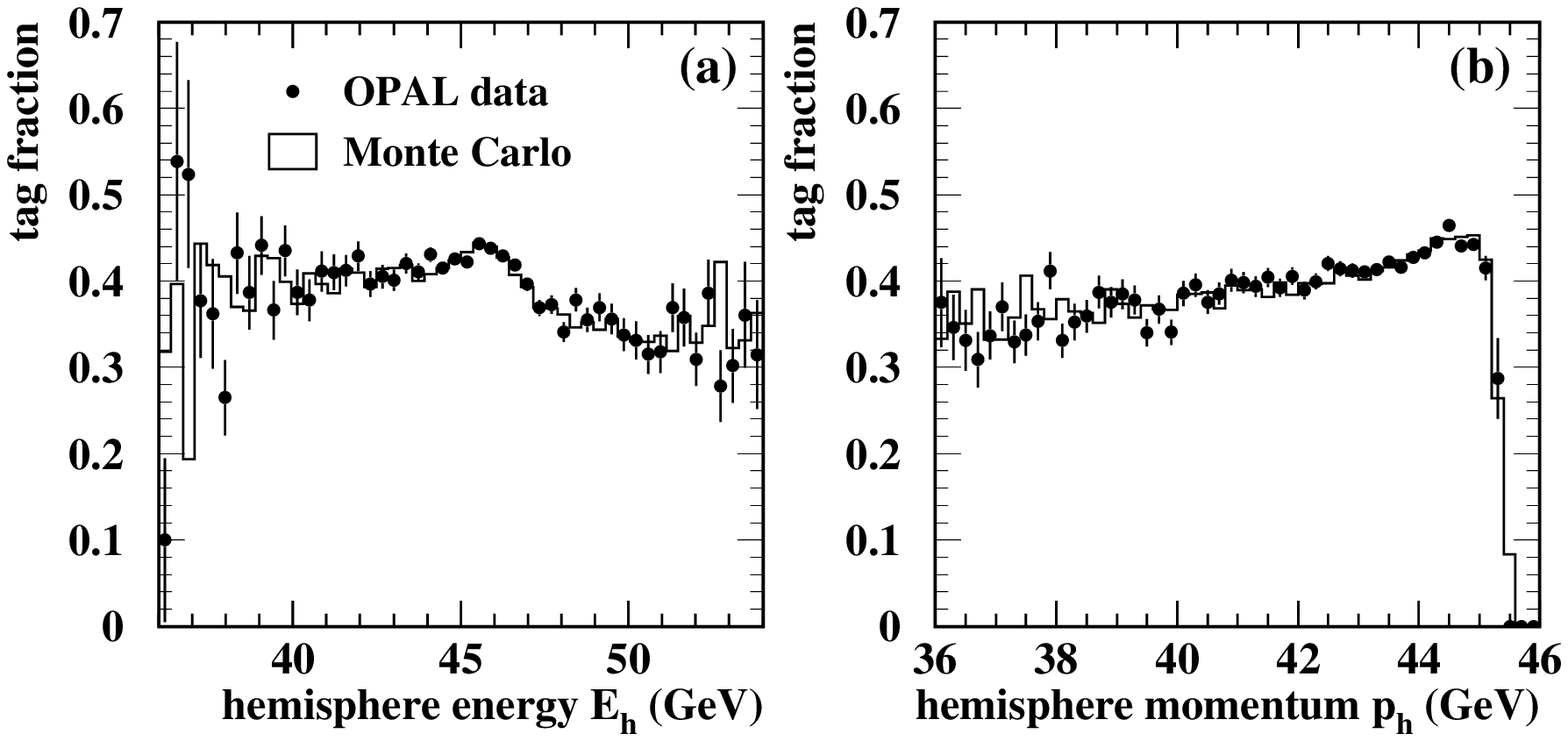}{f:cep2}{Fractions of tagged hemispheres in loose
  double tagged events as functions of (a) hemisphere energy \eh\ and
(b) hemisphere momentum \ph. The data are shown by the points and Monte Carlo
  by the histograms.}

\subsection{Geometrical Correlation}\label{ss:geoc}

The two b hadrons in a \bbbar\ event tend to be produced
back-to-back, so their decay products are likely to hit geometrically
opposite parts of the detector. This introduces an efficiency
correlation if the tagging efficiency is not directionally uniform.
The vertex tag efficiency depends on the polar angle
$\theta$ of the b hadron because multiple scattering degrades
the tracking resolution more 
as $\cos\theta$ increases. For some of the data, the
tagging efficiency also depends on $\phi$, 
due to inefficient regions of the silicon detector. The lepton tag 
efficiency depends on both $\theta$ and $\phi$, due to localised
regions of inefficiency caused by effects such as incomplete  muon
chamber coverage.

The size of the correlation from this effect
can be estimated from the data, by measuring the hemisphere tagging
probability as a function of the thrust axis polar angle $\theta_T$
and azimuthal angle $\phi_T$:
\begin{equation}\label{e:cbgeo2}
\cbgeom  = \frac{4\mean{f^+(\theta_T,\phi_T)f^-(\theta_T,\phi_T)}}
{\mean{f^+(\theta_T,\phi_T)+f^-(\theta_T,\phi_T)}
 \mean{f^+(\theta_T,\phi_T)+f^-(\theta_T,\phi_T)}}
\end{equation}
where $f^+$ ($f^-$) is the fraction of hemispheres in the $+z$ ($-z$)
direction that are tagged, and the averages are performed
over all values of  $\cos\theta_T$ and $\phi_T$.
The values of $f^+$ and $f^-$ are calculated in small bins of 
$\cos\theta_T$ and $\phi_T$.
As the tagged  samples are very pure in
\bbbar\ events, contributions from charm and light flavour events to
the correlations calculated from the tagged data are negligible.

The resulting geometrical correlations for the vertex, lepton and
combined tags  are given
for each year of the data in Table~\ref{t:geoc}. The values for the
different years are consistent, and  in
agreement with those derived from Monte Carlo simulation of the
detector configuration in each year.

\begin{table}
\centering

\renewcommand{\arraystretch}{1.3}
\begin{tabular}{ll|rrrr|r}\hline\hline
Tag & & \multicolumn{1}{c}{1992} & \multicolumn{1}{c}{1993} &
\multicolumn{1}{c}{1994} & \multicolumn{1}{c|}{1995} &
\multicolumn{1}{c}{Monte Carlo} \\ \hline
Vertex & $\cbgeom-1$ (\%) & 
$0.89\pm 0.14$ & $1.15\pm 0.17$ & $0.82\pm 0.09$ & $1.04\pm 0.13$ &
$0.88\pm 0.02$ \\
Lepton & $\cbgeom-1$ (\%) &
$1.02\pm 0.36$ & $0.54\pm 0.30$ & $0.33\pm 0.18$ & $0.12\pm 0.29$ &
$0.31\pm 0.03$ \\
Combined & $\cbgeom-1$ (\%) & 
$0.79\pm 0.12$ & $0.91\pm 0.11$ & $0.70\pm 0.06$ & $0.73\pm 0.09$ &
$0.71\pm 0.02$ \\
\hline
\end{tabular}
\caption{\label{t:geoc} Geometrical hemisphere efficiency correlations
  $\cbgeom-1$ and their statistical errors, 
  evaluated from each year of the data and from the Monte Carlo
  \bbbar\ sample.}
\end{table}

In order to calculate the overall correlation coefficient appropriate 
for each year of the data, the geometrical correlation in the Monte
Carlo sample was calculated using equation~\ref{e:cbgeo2}. These values
are also given in Table~\ref{t:geoc}. A correction:
\begin{equation}\label{e:cdelgeo}
\Delta\cb=\cbgeod-\cbgeomc
\end{equation}
was then applied to the overall Monte Carlo correlation to give the final
correlation value for each year of the data shown in Table~\ref{t:cbsum}.

The geometrical correlation is calculated using the thrust axis as an
estimator of the b hadron direction, assuming the two b hadrons to be
back to back. Final state gluon radiation can push the two b hadrons
away from the thrust axis direction, and towards regions of the
detector with different b tagging efficiency. The size of the true
geometrical correlation is therefore given not by
equation~\ref{e:cbgeo2} but by:
\begin{equation}\label{e:cbgeotr}
\cbgeom = \frac{\mean{\eb(\costhb,\phib)\eb(\costhbbar,\phibbar)}}
{\mean{\eb(\costhb,\phib)}\mean{\eb(\costhbbar,\phibbar)}} ,
\end{equation}
where \costhb, \phib\ and \costhbbar, \phibbar\ give the directions of
the two b hadrons, approximated by the thrust axis direction
in equation~\ref{e:cbgeo2}.
The error due to this approximation
was estimated by calculating $\Delta\cb$ using
equation~\ref{e:cdelgeo}, but with \cbgeod\ replaced by \cbgeomc\ calculated
from an independent  Monte Carlo sample with a geometrical correlation
twice as strong as that seen in
the data samples. The value of $\Delta\cb$ calculated using the thrust
axis direction was then compared with that calculated using the true 
b-hadron directions, evaluating the two \cbgeomc\ values using 
equation~\ref{e:cbgeotr}. The difference was $0.03\,\%$ for the vertex tag
alone and $0.02\,\%$ for the combined tag, which was taken as
the systematic error on the geometrical correlation correction.

\subsection{Primary Vertex Correlation}\label{ss:cpvtx}

In the previous analysis \cite{pr188}, the primary vertex position was
determined using all tracks in the event, and then used in the tagging
of each hemisphere. This led to a small but significant negative efficiency 
correlation. In this analysis, the primary vertex correlation has been
largely eliminated by determining the primary vertex separately for
each hemisphere (see Section~\ref{ss:pvtx}). However, the beam spot
constraint is still shared between the hemispheres, which may lead to
a small residual correlation.

This was investigated in the Monte Carlo by generating separate independent
beam spot constraints for each hemisphere, according to a Gaussian
distribution of the same width as the usual beam spot constraint, 
centred about the true event primary vertex position.
This procedure retains the constraint from the beam spot
but removes any correlation due to the shared position information.
The Monte Carlo tagging efficiency correlation using this modified 
constraint was found to be different by only $0.02\,\%$ from that with
the normal constraint, showing there is no significant primary vertex 
correlation from the shared beamspot. The full size of this shift is
taken as an additional systematic error on the overall correlation value.

\subsection{Correlation Simulation Systematics}\label{ss:csim}

The estimate of the overall size of the efficiency correlation from
the Monte Carlo relies on the detector simulation
and the modelling of the production and decay of b hadrons. The
following uncertainties were considered to estimate the systematic
error from these sources. The effect of each variation on the 
correlation values is given in Table~\ref{t:cbsum}.

\begin{description}
\item[Detector Resolution:] The detector resolution was varied using
  the method discussed in section~\ref{s:detsys}, varying the
  $r$-$\phi$ resolution parameter in the range 0.9--1.1. Since the effect seen
  is very small, additional studies varying the $r$-$z$ tracking
  resolution, tracking efficiency, silicon alignment and silicon hit
  matching efficiency were not carried out.

\item[Beam spot size:] The size of the LEP beam spot varied between
  each year of the data due to the differing operating conditions of the
  accelerator. The beam spot size was measured in each year
  using $\zb\rightarrow\mu^+\mu^-$ events, and was typically $140\rm\,\mu m$ in
  $x$, $10\rm\,\mu m$ in $y$, and $7\rm\,mm$ in $z$.
  In $x$ and $z$, the beam spot size also
  decreases gradually during the course of each fill, due to the
  operation of the LEP wiggler magnets. Residual uncertainties in the
  beam spot size are estimated to be $\pm 10\,\mu m$ in $x$, 
  $\pm 5\,\mu m$ in $y$ and $\pm 1\rm\,mm$ in $z$. The effective 
  beam spot constraint in Monte Carlo events was varied by these
  amounts to assess the resulting uncertainty in the correlation 
  coefficients.

\item[b quark fragmentation:] The b quark fragmentation was varied in
  the Monte Carlo by  applying a weight to each simulated event using
  the fragmentation functions discussed in Section~\ref{ss:esimi}, so as
  to vary the average scaled energy \meanxe\ of the weakly decaying b hadrons
  by $\pm 0.008$. This variation represents the accuracy of \meanxe\
  measured by LEP experiments \cite{bfrag,bdfrag}.

\item[b hadron lifetime:] The lifetimes of the b hadrons were varied 
simultaneously by $\pm 0.05\rm\,ps$ using a weighting method. The size
of the variation was chosen to be larger than the accuracy of the
world average b-lifetime \cite{pdg98} to allow for the uncertainties
due to different efficiencies for different b hadron species.

\item[b hadron charged decay multiplicity:] The average charged decay
  multiplicity of the b hadrons was varied by $\pm 0.35$ using a
  weighting method. The size of the variation reflects the accuracy of
  the measurements by DELPHI and OPAL \cite{bmult}.

\end{description}
The resulting variations in the correlation coefficients are all
rather small, and much smaller than the dominant uncertainties caused by
momentum correlation effects.

\subsection{Correlation Completeness Test}\label{ss:compl}

\epostfig{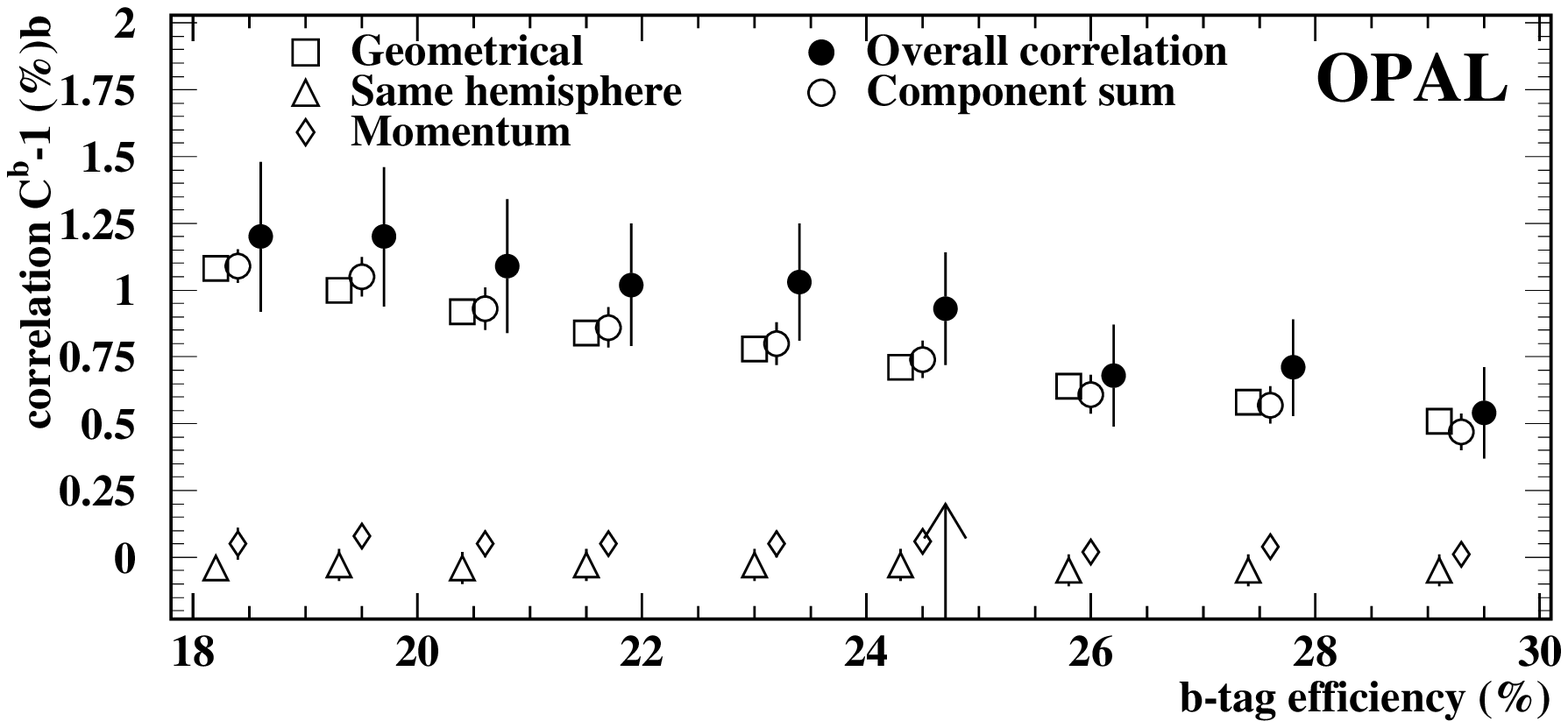}{f:corlvar}{Comparison of the true overall
  tagging efficiency correlation and the various components for the
  combined vertex and lepton tag, as the b-tagging efficiency is
  varied by changing the vertex tag cut. The individual components are
  slightly displaced along the $x$-axis for clarity, and the analysis
  cut is shown by the arrow.}

The sizes of the various components of the correlation in Monte Carlo,
together with their associated statistical errors,
are given in Table~\ref{t:compl}. 
The evolution of the various components as the b-tagging efficiency is
varied is shown in Figure~\ref{f:corlvar}.
In the absence of
interdependence between the correlation components, the values of
$\cb-1$ from each component can be added to give the overall
correlation. The sum of the components in Table~\ref{t:compl} is
compatible with the overall correlation evaluated from the ratio of
the event tagging efficiency to the product of the hemisphere tagging
efficiencies, for both vertex and combined
tags. This strongly suggests that no large correlation contribution has
been omitted from the systematic error evaluation, though the
precision of this test is limited by the overall Monte Carlo
statistics and by interdependence between the different components.
In the previous analysis \cite{pr188}, an additional systematic 
error was included to cover such interdependence, since the overall
correlation was determined by simply adding the components.
No such error is included here,
since the Monte Carlo is used to determine the value of the
correlation directly, including any interdependence.

\begin{table}
\centering

\begin{tabular}{l|rr}\hline\hline
Correlation $\cb-1$ (\%) & Vertex & Combined \\
\hline
Same hemisphere events &  $0.02\pm 0.02$ & $-0.03\pm 0.02$ \\
Momentum correlation    &  $0.04\pm 0.05$ &  $0.06\pm 0.03$ \\
Geometrical correlation &  $0.88\pm 0.02$ &  $0.71\pm 0.02$ \\
\hline
Component sum           &  $0.94\pm 0.06$ &  $0.74\pm 0.04$ \\
\hline
Overall correlation     &  $0.83\pm 0.20$ &  $0.93\pm 0.17$ \\
\hline
\end{tabular}
\caption{\label{t:compl} Components of the hemisphere efficiency
  correlation $\cb-1$ in \bbbar\ Monte Carlo events, together with
  their sum and the overall correlation in the Monte Carlo
  sample. Only statistical errors are included. }
\end{table}

These correlation component sums have been made using the \cbpp\
estimate from equation~\ref{e:cbpp} for the momentum
correlation. If the energy and momentum conservation \cbep\ 
estimate from equation~\ref{e:cbep} is
used instead, the component sums are $(0.63\pm 0.05)\,\%$ and 
$(0.54\pm 0.04)\,\%$ for the vertex and combined tags, still
compatible with the overall Monte Carlo correlation. It should be
emphasised that the central value used for the correlation does not
depend on which method is used to evaluate the momentum correlation
component.

\section{Summary and  Conclusion}\label{s:concl}

The fraction \rb\ of \ztobb\ events in hadronic \zb\ decays  was
measured using data collected by OPAL between 1992 and 1995 using tags
based on displaced secondary vertices and identified leptons, 
giving the result
\[
\rb =\rbval \pm \rbstat \pm \rbsyst
\]
where the first error is statistical and the second systematic. The
systematic error does not include the effects of varying \rc\ from its
Standard Model expectation. The result depends on \rc\ as follows:
\[
\frac{\Delta\rb}{\rb}=\delrbrc\frac{\Delta\rc}{\rc}
\]
where $\Delta\rc$ is the deviation of \rc\ from the value 0.172
predicted by the Standard Model. 

This result is in agreement with
and  supersedes our previous measurement \cite{pr188}. It is also in
agreement with other recent measurements at LEP \cite{alephonetag,otherrblep}
and SLC \cite{otherrbslc}, and with the
Standard Model prediction of $\rbtheo\pm\rbterr$ \cite{ZFITTER}.

\section*{Acknowledgements}

We particularly wish to thank the SL Division for the efficient operation
of the LEP accelerator at all energies
 and for their continuing close cooperation with
our experimental group.  We thank our colleagues from CEA, DAPNIA/SPP,
CE-Saclay for their efforts over the years on the time-of-flight and trigger
systems which we continue to use.  In addition to the support staff at our own
institutions we are pleased to acknowledge the  \\
Department of Energy, USA, \\
National Science Foundation, USA, \\
Particle Physics and Astronomy Research Council, UK, \\
Natural Sciences and Engineering Research Council, Canada, \\
Israel Science Foundation, administered by the Israel
Academy of Science and Humanities, \\
Minerva Gesellschaft, \\
Benoziyo Center for High Energy Physics,\\
Japanese Ministry of Education, Science and Culture (the
Monbusho) and a grant under the Monbusho International
Science Research Program,\\
Japanese Society for the Promotion of Science (JSPS),\\
German Israeli Bi-national Science Foundation (GIF), \\
Bundesministerium f\"ur Bildung, Wissenschaft,
Forschung und Technologie, Germany, \\
National Research Council of Canada, \\
Research Corporation, USA,\\
Hungarian Foundation for Scientific Research, OTKA T-016660, 
T023793 and OTKA F-023259.\\


\newpage

\begin{thebibliography}{99}
\bibitem{bamert}
See for example:~P.~Bamert~\etal, \PRD{54}{1996}{4275}, and references
therein.

\bibitem{ZFITTER}
    D.~Bardin~\etal, CERN-TH 6443/92, May 1992. \\
We use ZFITTER version 5.0 with default parameters, and the following
inputs: $m_{\rm Z}=91.1867\pm 0.0020\rm\,GeV$, 
$m_{\rm t}=175.6\pm 5.5\rm\,GeV$, 
$\alpha(m_{\rm Z})=1/128.909$, $\alpha_s(m_{\rm Z})=0.120\pm 0.003$ and
$m_{\rm H}=115^{+116}_{-38}\rm\,GeV$.

\bibitem{theogcc}
M.H.~Seymour, \NPB{436}{1995}{163}.

\bibitem{pr188}
    OPAL collaboration, K.~Ackerstaff~\etal, \ZPC{74}{1997}{1}.

\bibitem{opalsi3d}
P.P.~Allport~\etal, \NIM{A346}{1994}{476}.

\bibitem{opaldet}
OPAL collaboration, K.~Ahmet~\etal, \NIM{A305}{1991}{275}; \\
P.P.~Allport~\etal, \NIM{A324}{1993}{34}.

\bibitem{jetcone}
OPAL collaboration, R.~Akers~\etal, \ZPC{63}{1994}{197}.

\bibitem{jetset}
T.~Sj\"ostrand, \CPC{82}{1994}{74}.

\bibitem{jetsetopt}
OPAL collaboration, G.~Alexander~\etal, \ZPC{69}{1996}{543}.

\bibitem{fpeter}
C.~Peterson, D.~Schlatter, I.~Schmitt and P.~Zerwas, \PRD{27}{1983}{105}.

\bibitem{gopal}
J.~Allison~\etal, \NIM{A317}{1992}{47}.

\bibitem{alephonetag}
ALEPH collaboration, R.~Barate~\etal, \PLB{401}{1997}{150}.

\bibitem{beamspot}
OPAL collaboration, P.D.~Acton \etal , \ZPC{59}{1993}{183};\\
OPAL collaboration, R.~Akers \etal , \PLB{338}{1994}{497}.

\bibitem{jetnet}
The neural networks were trained using JETNET 3: \\
C.~Peterson, T.~R\"ognvaldsson and L.~L\"onnblad, \CPC{81}{1994}{185}.

\bibitem{elecid}
OPAL collaboration, R.~Akers~\etal, \ZPC{70}{1996}{357}.

\bibitem{muonid}
OPAL collaboration, P.D.~Acton~\etal, \ZPC{58}{1993}{523}.

\bibitem{pi0prod}
OPAL collaboration, K.~Ackerstaff~\etal, CERN-EP/98-054, accepted by
Eur.\ Phys.\ J.\ C.

\bibitem{bdfrag}
ALEPH Collaboration, D.~Buskulic~\etal, \ZPC{62}{1994}{179}.

\bibitem{dboth}
DELPHI collaboration, P.~Abreu~\etal, \ZPC{59}{1993}{533}, erratum in
\ZPC{65}{1995}{709};\\
OPAL collaboration, G.~Alexander~\etal, \ZPC{72}{1996}{1}.

\bibitem{dfrag}
ALEPH collaboration, D.~Buskulic~\etal, \ZPC{62}{1994}{1};\\
OPAL collaboration, R.~Akers~\etal, \ZPC{67}{1995}{27}.

\bibitem{hfew}
The LEP collaborations, ALEPH, DELPHI, L3 and OPAL,
\NIM{A378}{1996}{101}.\\
Updated averages are described in `Presentation of LEP Electroweak
Heavy Flavour Results for Summer 1998 Conferences', LEPHF 98-01
(see {\tt http://www.cern.ch/LEPEWWG/heavy/} ).

\bibitem{fcolspil}
P.~Collins and T.~Spiller, \JPH{G11}{1985}{1289}.

\bibitem{fkart}
V.G.~Kartvelishvili, A.K.~Likhoded and V.A.~Petrov, \PLB{78}{1978}{615}.

\bibitem{flund}
B.~Anderson, G.~Gustafson and B.~S\"oderberg, \ZPC{20}{1983}{317}.

\bibitem{dprod}
ALEPH collaboration, D.~Buskulic~\etal, \ZPC{69}{1996}{585}.

\bibitem{pdg98}
Particle Data Group, C.~Caso~\etal, \EPJ{3}{1998}{1}.

\bibitem{mark3}
MARK III collaboration, D.~Coffman~\etal, \PLB{263}{1991}{135}.

\bibitem{pdg96}
Particle Data Group, R.M.~Barnett~\etal, \PRD{54}{1996}{1}.

\bibitem{clargus}
ARGUS collaboration, H.~Albrecht~\etal, \PLB{278}{1992}{202}.

\bibitem{accm}
G.~Altarelli~\etal, \NPB{208}{1982}{365}.

\bibitem{delco}
DELCO collaboration, W.~Bacino~\etal, \PRL{43}{1979}{1073}.

\bibitem{mark3lept}
MARKIII collaboration, R.M.~Baltrusaitis~\etal, \PRL{54}{1985}{1976}.

\bibitem{opalgcc}
OPAL collaboration, R.~Akers~\etal, \ZPC{67}{1995}{27};\\
OPAL collaboration, R.~Akers~\etal, \PLB{353}{1995}{595}.

\bibitem{adgbb}
ALEPH collaboration, R.~Barate~\etal, `A measurement of the gluon
splitting rate into \bbbar\ pairs in hadronic Z decays',
CERN-EP/98-103, accepted by Phys.\ Lett.\ B; \\
DELPHI collaboration, P.~Abreu~\etal, \PLB{405}{1997}{202}.

\bibitem{opalklh}
OPAL collaboration, R.~Akers~\etal, \ZPC{67}{1995}{389}; \\
OPAL collaboration, G.~Alexander~\etal, \ZPC{73}{1997}{569}; \\
OPAL collaboration, G.~Alexander~\etal, \ZPC{73}{1997}{587}.

\bibitem{herwig}
G.~Marchesini~\etal, \CPC{67}{1992}{465}.

\bibitem{nasoncb}
P.~Nason and C.~Oleari, \PLB{407}{1997}{57}.

\bibitem{gcecode}
OPAL collaboration, M.~Akrawy~\etal, \PLB{253}{1990}{511}.

\bibitem{bfrag}
ALEPH Collaboration, D.~Buskulic~\etal, \PLB{357}{1995}{699};\\
DELPHI Collaboration, P.~Abreu~\etal, \ZPC{66}{1995}{323};\\
OPAL Collaboration, R.~Akers~\etal, \ZPC{60}{1993}{199};\\
OPAL Collaboration, G.~Alexander~\etal, \PLB{364}{1995}{93}.

\bibitem{bmult}
DELPHI Collaboration, P.~Abreu~\etal, \PLB{347}{1995}{447};\\
OPAL Collaboration, R.~Akers~\etal, \ZPC{61}{1994}{209}.

\bibitem{otherrblep}
ALEPH collaboration, R.~Barate~\etal, \PLB{401}{1997}{163};\\
DELPHI collaboration, P.~Abreu~\etal, \ZPC{70}{1996}{531}.

\bibitem{otherrbslc}
SLD collaboration, K.~Abe~\etal, \PRL{80}{1998}{660}.

\end{thebibliography}
\end{document}